\documentclass[prd,nofootinbib,tightenlines,twocolumn]{revtex4-1} 
\usepackage{bm,dcolumn,amsmath,graphicx,amsfonts,amssymb}

\usepackage{hyperref} 
\usepackage{xcolor}
\usepackage{float}
\definecolor{cite}{rgb}{0.,0.,0.85}   
\hypersetup{colorlinks,linkcolor={cite},citecolor={cite},urlcolor={cite}}
\usepackage{soul}

\newcommand{\abs}[1]{\ensuremath{\left |#1\right |}}
\newcommand{\av}[1]{\ensuremath{\langle #1\rangle}}	

\renewcommand{\v}[1]{\ensuremath{\boldsymbol{#1}}}		
\newcommand{\units}[1]{\ensuremath{ \,\mathrm{#1}} }

\DeclareMathOperator\Prob{Prob}
\DeclareMathOperator\var{Var}

\def\h{\ensuremath{\hbar}}

\renewcommand{\a}{\ensuremath{\alpha}}

\renewcommand{\t}{\ensuremath{\tau}} 
\renewcommand{\k}{\ensuremath{\kappa}}

\def\CT{\ensuremath{{\cal T}}}
\def\L{\ensuremath{{\cal L}}}

\newcommand{\un}[1]{\ensuremath{\,{\rm{#1}}}} 

\usepackage[normalem]{ulem}
\definecolor{newc}{rgb}{0.,0.,0.5}
\definecolor{newerc}{HTML}{03B4C8}

\usepackage{url}

\renewcommand{\bar}[1]{\overline{#1}}
\newcommand{\Geff}{\ensuremath{\Gamma_{\rm eff}}}

\newcommand{\ND}{\ensuremath{N_{\rm D}}}

\newcommand{\be}{\begin{equation}}
\newcommand{\ee}{\end{equation}}

\begin{document} 

\title{Applying the matched-filter technique to the search for dark matter transients with networks of quantum sensors}

\author{G. Panelli}
	\affiliation{Department of Physics, University of Nevada, Reno, 89557, USA}
\author{B. M. Roberts}
\altaffiliation[Present address:~]{School of Mathematics and Physics, the University of Queensland, Brisbane, QLD 4072, Australia}
    \affiliation{Department of Physics, University of Nevada, Reno, 89557, USA} 
\author{A. Derevianko}
	\affiliation{Department of Physics, University of Nevada, Reno, 89557, USA}

\date{ \today }  

\begin{abstract}
There are several networks of precision quantum sensors in existence, including networks of atomic clocks, magnetometers, and gravitational wave detectors. These networks can be re-purposed for searches of exotic physics, such as direct dark matter searches. Here we explore a detection strategy for macroscopic dark matter objects with such networks using  the matched-filter technique. Such ``clumpy'' dark matter objects would register as transients sweeping through the network at galactic velocities. 
As a specific example, we consider a network of atomic clocks aboard the Global Positioning System (GPS) satellites. We apply the matched-filter technique  to simulated GPS atomic clock data and study its utility and performance.  The analysis and the developed methodology have a wide applicability to other networks of quantum sensors. 
\end{abstract}

\maketitle

 \section{Introduction}\label{sec:intro}

Astrophysical observations on the galactic scale indicate that dark matter (DM) constitutes 85$\%$ of all matter in the universe, leaving only 15$\%$ to ordinary matter~\cite{Bertone2005}. These galactic scale observations (e.g., rotation curves, cosmic microwave background, gravitational lensing) characterize solely the gravitational interaction of dark matter and ordinary matter, leaving  the microscopic composition  of DM and its non-gravitational interactions with standard model particles and fields  a mystery. Among the current DM searches, weakly interacting massive particles (WIMPs) have been the primary target with no success to date, thereby partially motivating alternative DM candidates~\cite{Bertone2005}.  One such alternative is ultralight fields where the DM candidate may take the form of a macroscopic object coherent over large scales. Such objects would exert gentle minute perturbations detectable by quantum sensors~\cite{Degen2017-RMP-quantum-sensing}. Quantum sensors are typically well protected from environmental perturbations which makes them uniquely sensitive to new physics. Examples include atomic clocks, magnetometers, atom interferometers, and  microwave and optical cavities. The particular hypothesized form of coupling of DM fields to baryonic matter determines the type of the sensor to be used in the DM search~\cite{Safronova2018-RMP}.

Here we focus exclusively on ultralight fields. In general, one may consider free non-interacting and self-interacting fields. We are interested in models with self-interaction that can form macroscopic dark-matter objects, such as topological defects or Q-balls~\cite{Vilenkin:1984ib,COLEMAN1985263,Kusenko2001,LEE1989}.  
Interactions of such DM constituents with standard model (SM) particles and fields can induce variations in fundamental constants of nature.
Such variations may be detected by observing atomic frequencies in atomic clocks~\cite{LIU2017, HUNTEMANN14, GODUN2014}. 
As the DM constituents sweep through a clock, the variation would register as a transient perturbation. Geographically-distributed networks can resolve the velocities of the transients and provide a powerful vetoing of the potential events as the sweep velocity must be consistent with standard halo model priors.  These ideas form the basis of dark matter searches with distributed  networks~\cite{POSPELOV2013,Derevianko2014b,Roberts2017,Roberts2018b,Wcislo2018}.

Our GPS.DM group focuses on searching for DM-induced transients in GPS atomic clock data~\cite{Derevianko2014b,Roberts2017}.
Here the advantage is the public availability of nearly two decades of archival data enabling relatively inexpensive data mining.
A dark matter signature would consist of a correlated propagation of atomic clock perturbations through the GPS constellation at galactic ($\sim 300 \,  \mathrm{km/s}$) velocities. 
Previously, our GPS.DM collaboration performed an analysis of the archival GPS data in search for domain walls (a particular type of topological defect)~\cite{Roberts2017}.  Although no DM signatures were found, new limits were placed on certain DM couplings to atoms that were several orders of magnitude more stringent than prior astrophysical limits.  The original search~\cite{Roberts2017} focused on finding large DM signals well above the instrument noise. 
In Ref.~\cite{Roberts2018b}, we have shown that the application of more sophisticated Bayesian search techniques can extend the discovery reach by several orders of magnitude both in terms of sensitivity and size/geometry of the DM objects. Here we study the performance of the matched-filter technique (MFT) as an alternative frequentist search method. In addition, we develop analytic results for idealized network of white-noise sensors with cross-node correlation.

The MFT is a relatively ubiquitous technique, utilized, for example, by the Laser Interferometer Gravitational Wave Observatory (LIGO) in gravitational wave detection (see, e.g.~\cite{RomanoCornish2017}). 
It is also used in a variety of applications, such as astrophysics~\cite{Dong2008}, geophysics~\cite{Shearer1994}, and searches for exotic physics~\cite{Auger2012}.
Of a special interest to us is the performance of the MFT for large networks, as the GPS.DM sensor array may include over a hundred instruments if we take into account all GNSS constellations and terrestrial clocks. 
A technical complication in applying the MFT to these networks is that the noise is  correlated between spatially separated nodes of the network. Understanding the consequences of this cross-node correlation is of importance to our analysis.

The structure of this paper is as follows. Sec.~\ref{sec:mft} reviews the MFT for data analysis and previous applications.  Sec.~\ref{Sec:Desiderata} formalizes requirements for a network detector of clumpy DM. 
Sec.~\ref{sec:theory} reviews the relevant theory and how the GPS network can be re-purposed for DM searches of this sort. 
A summary of processing the GPS data is provided in Sec.~\ref{sec:data}. 
In Sec.~\ref{sec:method}, we describe our methodology, including the formulation of our detection statistic and the search algorithm. 
The detection threshold, detection probability and parameter estimation capabilities are provided in 
Secs.~\ref{sec:thresh}, \ref{sec:dp}, and \ref{sec:param}, respectively.
Lastly, the projected discovery reach for this method is provided in Sec.~\ref{sec:reach} and Sec~\ref{Sec:Conclusions} draws conclusions.
The paper also contains  appendices where we discuss  network covariance matrix and its inversion, inverse transform sampling, and present supporting derivations. Since the intended audience includes both atomic and particle physics communities,  we restore $\hbar$ and $c$ in the formulas in favor of using natural or atomic units. We also use the rationalized Heaviside-Lorentz units for electromagnetism.

\section{The matched-filter technique} \label{sec:mft}
The matched-filter technique is often used to search for hidden signals within data streams in cases where the signal's ``shape'' is known but the signal's strength is not. In this case, the general shape can be compared to the data stream to search for an underlying correlation that is not immediately evident to the un-aided eye. 
The matched-filter itself is the best estimate of the unknown signal strength by employing an optimal filter. 
In the most general sense, an optimal filter is a particular combination of data that optimizes a quantity deemed to be significant, usually relating to signal detection within data sets~\cite{RomanoCornish2017}.
Usual quantities of interest include the detection probability for a given signal strength and the signal-to-noise ratio (SNR), though many other application-specific statistics can be devised.
 
Since the MFT requires a predefined signal shape, this approach cannot be used for un-modeled signals. 
When a hypothesized signal signature is able to be modeled, there often exist many candidate shapes along with the unknown signal strength. 
Thus, one forms a collection of signal shape templates that approximately spans the range of possible shapes.
One could think of the MFT as a technique that maximizes an overlap between the templates and the data stream. 
This maximization is done with the help of a matched-filter statistic, such as a SNR. 
However, it is not usually the value of the SNR alone that determines the level of overlap, but rather the value of the SNR compared to a threshold.
Additionally, one of the most promising aspects of the MFT is the ability to align the shape of the detected signal with the template that produced an SNR above the detection threshold, leading to immediate signal parameter estimation. 
 
The efficacy of the MFT depends on how well one can distinguish a weak signal from intrinsic device noise. 
Thus, a network of devices can offer a better sensitivity and higher confidence in the event of a positive detection since all network sensors would experience a signal from the same event.\footnote{This is technically only true for the domain-wall DM object as domain walls sweep through all the network nodes. Other DM objects such as monopoles may only affect a fraction of the nodes.} 
If a hypothesized signal shape can be modeled for a network of devices, the MFT becomes a powerful tool for weak signal detection and signal parameter estimation.

\subsection{Examples of the MFT in practice}
Perhaps the most well-known application of the MFT comes from the gravitational wave detection by LIGO (see e.g.,~\cite{Abbott2012}). A detailed outline of search techniques and matched-filtering is provided in Ref.~\cite{RomanoCornish2017} and guided much of our development of the method discussed in this paper. 
However, LIGO's use of the MFT involved a small network of devices that exhibit uncorrelated noise. 
The black hole merger gravitational wave detection from 2015, for instance, used only two spatially separated interferometers in the waveform template matching analysis~\cite{Abbott2012}.
Another previous application of the MFT used 15-20 station from the International Deployment of Accelerometers (IDA), where geophysicists were been able to identify previously undetected global seismic events in archival IDA data~\cite{Shearer1994}.
The method has also found use in galaxy cluster identification~\cite{Dong2008} and in the search for neutrino-less beta decay~\cite{Auger2012}. 

All of these examples consider networks of devices far smaller than that available ($\sim 100$) to our DM search. 
Furthermore, an essential feature of our GPS.DM network search is  a cross-node correlated noise (due to a reference clock common to all the nodes, see Sec.~\ref{sec:data}), which, to the best of our knowledge, has not been addressed in the literature.  A related theoretical development for ultralight (non self-interacting) DM fields is presented in Ref.~\cite{Derevianko2016a}.  There a quantum sensor network was considered for detecting DM waves and a SNR statistic was developed in the frequency domain. By contrast, here we focus on network detection of dark matter transients and develop a SNR statistic in the time domain.

\section{Network desiderata}
\label{Sec:Desiderata}
Here we are devising a strategy for detecting macroscopic DM objects that sweep through a distributed network of $\ND$ sensors. We use the words ``sensor'' and  ``node'' interchangeably. In particular, a single geographical location may host several instruments, yet each individual sensor is referred to as a distinct node for our purposes. We will assume that DM objects interact (non-gravitationally) with the instruments and the interaction only occurs when the bulk of the DM object overlaps with a sensor. 

There are several criteria for such networks: 
\begin{enumerate}
\item[(i)]{The network should be sufficiently dense so that the finite-size DM object can overlap with at least several geographically-distinct nodes. }
\item[(ii)]{The network volume should be sufficiently large to increase the rate of encounters with DM objects. }
\item[(iii)]{ As per the standard halo model (SHM)~\cite{BOVY2012} the DM objects sweep through the network at galactic velocities ($v_g \sim 300 \,\mathrm{km/s}$),
the sampling rate should be sufficiently high to enable tracking the propagation of the DM object through the network. The tracking enables reconstructing the geometry of the encounter.  }
\item[(iv)]{Although not necessary, it is desired that the encounters of DM objects with the network are sufficiently rare, so that only a single DM object interacts with the network at any given time.}
\end{enumerate} 

While most of these requirements are apparent, criterion (iii) deserves further discussion.
For example, while a setup~\cite{Wcislo2016}  of two co-located clocks with a shared optical cavity can be considered as a rudimentary two-node network, such network does not satisfy criterion (iii) since even if both clocks were to register a DM signal, there would be no galactic velocity/direction signature to support the signal's DM origin. Thus, such low-sampling rate networks can only be used to constrain couplings to the DM sector.

\section{Clumpy dark matter models}\label{sec:theory}

Stable macroscopic objects  may form from ultralight DM fields due to their self-interaction in the dark sector~\cite{SIKIVIE1982, WH1989, VOLENKIN1994, DURRER20021, FRIEDLAND2003, AVELINO2008}. 
Topological defects (TDs) are an example of such macroscopic ``clumpy'' DM objects, though they can also contribute to dark energy depending on their cosmological fluid equation of state~\cite{Roberts2018b}. 
Monopoles (0D), strings (1D), and domain walls (2D) are all examples of TDs of various dimensionalities.  
Other examples of macroscopic DM candidates include $Q$-balls~\cite{COLEMAN1985263, Kusenko2001, LEE1989}, solitons~\cite{MARSH15, Schive2014}, and axion stars~\cite{HOGAN1988, KOLB1993}. 
A special case of clumpy DM are DM ``blobs''~\cite{Grabowska2018} --  particle-like DM objects sourcing long-range Yukawa-type interactions with the SM sector.  

For concreteness, we focus on topological defects. Inside the defect, the amplitude of the DM field $A$ and the average energy density of the defect is related by $\rho_{\rm inside}= A^2/(\h c \, d^2)$, where $d$ is the width or spatial scope of the defect (we use the convention where the field has units of energy).
The DM object width $d$ is treated as a free observational parameter and, for TD models, may be linked to the mass of the DM field particles $m_{\phi}$ through the healing length which is on the order of the Compton wavelength $d=\hbar/(m_\phi c)$.
Further, the local DM energy density $\rho_{\rm DM}$ may be linked to $d$ and $A$ by assuming that these objects saturate the local DM energy density,  
\begin{equation}
    A^2 = (\hbar c) \, \rho_{\rm DM} d^2 \frac{\CT}{\tau_{\text{avg}}} \, ,
\end{equation}
where $\tau_{\text{avg}} \sim d/v_g$ is the average duration of crossing through a point-like instrument and $\CT$ is the average time between subsequent encounters of the DM objects with the device~\cite{Roberts2017}.

As for the non-gravitational interactions, to be specific and consistent with our earlier work~\cite{Roberts2018b}, we assume the quadratic scalar portal,
\begin{equation}
\label{eq:scalarPortal}
-\L_\text{int}=  \left(
{\Gamma_f}{m_f c^2 \bar \psi_f \psi_f   } 
 +\Gamma_\alpha\frac{F_{\mu\eta}^2 }{4}
+ \,\ldots \right) \phi \phi^*  \, ,
\end{equation}
where $m_f$ are the fermion masses, 
$\phi$ is the scalar DM field (measured in units of energy), 
$\Gamma_{X}$ are coupling constants that quantify the DM interaction strength, and
$\psi_f$ and $F_{\mu\eta}$
are the SM fermion fields and the electromagnetic Faraday tensor, respectively. 
The SM fermions $f$ in the above equation are summed over implicitly. 
Such interactions appear naturally for DM fields possessing either $Z_2$ or $U(1)$ intrinsic symmetries.
The  above Lagrangian leads to 
an effective redefinition of fundamental masses and coupling constants,
\begin{align}
\label{eq:varalpha}
\alpha^{\rm eff}(\v{r},t) &= \left[1+ \Gamma_{\alpha}\,{|\phi(\v{r},t)|^2}\right]  {\alpha },\\
m_{f}^{\rm eff}(\v{r},t) &= \left[1+ \Gamma_{f}\,{|\phi(\v{r},t)|^2}\right] {m_{f}},
\label{eq:varqcd}
\end{align}
where $m_f$ are the fermion (electron/proton $m_{e/p}$ and light quark $m_q\equiv[m_u+m_d]/2$) masses and $\alpha\approx1/137$ is the electromagnetic fine-structure constant.
The coupling constants $\Gamma$ have units of  $[{\rm Energy}]^{-2}$ and we also define the effective energy scales as $\Lambda_X \equiv 1/\sqrt{\abs{\Gamma_X}}$ with $X=\alpha,\,m_{e/p},\,m_q$ to aid the comparison with previous literature.

The observable atomic frequency shift induced by the DM objects can be linked to the transient variation of fundamental constants, and thus the DM field parameter, from Eqs.~(\ref{eq:varalpha}) and (\ref{eq:varqcd}). 
For a particular clock transition,
\begin{equation}
\label{eq:variation}
 \frac{\delta \omega(\v{r}, t)}{ \omega_0} 
= \sum_X \k_X\Gamma_X{|\phi(\v{r},t)|^2} \equiv \Geff\,{|\phi(\v{r},t)|^2}, 
\end{equation}
where $\omega_0$ is the nominal clock frequency, $X$ runs over relevant fundamental constants, and $\k_X$ are dimensionless sensitivity coefficients.
For convenience, we introduced the effective constant, $\Gamma_{\rm eff} \equiv \sum_X \kappa_X \Gamma_X$, which depends on the specific clock. 

The effective coupling constants for the GPS network microwave atomic clocks (Rb, Cs and H) read (using computations~\cite{DZUBA2003, FLAMBAUM20062}, see \cite{Roberts2018b} for details, and Ref.~\cite{Savalle2019-DAMNED} for illucidating the underlying logic)
\begin{align}
\label{eq:Crb}
\Geff({\rm ^{87}Rb})&={4.34}\,{\Gamma_\a}-{0.069}\,{\Gamma_{m_q}}+{2}\,{\Gamma_{m_e}},
\\
\label{eq:Ccs}
\Geff({\rm ^{133}Cs})&={4.83}\,{\Gamma_\a}-{0.048}\,{\Gamma_{m_q}}+{2}\,{\Gamma_{m_e}},
\\
\label{eq:Ch}
\Geff({\rm ^{1}H})&={4}\,{\Gamma_\alpha}-{0.150}\,{\Gamma_{m_q}}+{2}\,{\Gamma_{m_e}}.
\end{align}

Although linear combinations of the coupling constants  differ for each type of clock in the GPS network, individual coupling constants $\Gamma_X$ (or, equivalently, individual energy scales $\Lambda_X$) can be obtained by combining the results from different clock types. 
Laboratory optical Sr clock has provided the most stringent constrains on $\Lambda_{\a}$ for specific regions of the $(d,\CT)$ parameter space (see, e.g.~\cite{Wcislo2016}).
More recently, new constraints have been placed on $\Lambda_{\a}, \Lambda_{m _e}$, and $\Lambda_{m_q}$ by our GPS.DM collaboration~\cite{Roberts2017} and on $\Lambda_{\a}$ by a global network of optical laboratory clocks~\cite{Wcislo-clock-network-2018}. These two papers reported null results for domain wall searches.

\subsection{Thin domain walls}
In this paper we will focus on a specific type of DM signal -- ``thin'' domain walls.
While retaining the main features of more complicated signals from other types of DM ``clumps'', this signal offers a sufficiently simple, analytically treatable signature. Domain wall-like signatures can appear naturally in the context of bubbles, i.e., domain walls closed on themselves~\cite{Roberts2017-GPS-DM}. Locally, one can neglect the bubble curvature as long as the bubble radius is much larger than the spatial extent of the sensor network. Another example are Q-balls  that couple to SM fields through derivative couplings, $\phi \phi^* \rightarrow \partial_\mu \phi \partial^\mu \phi^*$ in Eq.(\ref{eq:scalarPortal}). Q-balls are spherically symmetric objects with a nearly flat density profile in the bulk. Thus the dominant part of the interaction would occur at the Q-ball walls. Again one needs to require the radii of these objects to be much larger then the network size. 
Since bubbles and Q-balls are spherically symmetric, gravitationally interacting ensembles of these DM objects are a subject to the equation of state for pressureless cosmological fluid as required by the $\Lambda$CDM paradigm.   

We distinguish between ``thin'' and ``thick'' walls based on the sampling rate, which is finite for any realistic device.
If the interaction time with the device $d/v_g$ is shorter than the sampling interval $\tau_0$, the exact arrival time of the DM clump is not resolved, and neither its shape. Thus the DM object is ``thin'' for observational purposes if its size $d \ll v_g \tau_0$. For domain walls, strictly speaking, the relevant velocity is its component normal to the wall, $v_\perp$. 

For the GPS sampling interval of  $\tau_0 = 30\,\un{s}$, the above arguments translate into domain walls of thicknesses below the Earth size,  
 $d\ll  300\un{km}\un{s}^{-1}\times 30\un{s} \approx 10^4 \un{km}$ .
Any domain wall with a thickness larger than this value is characterized as ``thick''  and is discussed in more detail in Ref.~\cite{Roberts2018b}.

The thin wall network signature is formalized in Sec.~\ref{sec:mc}. For thin walls, the value of the effective coupling relates to the maximum DM-induced accumulated clock phase (time) signal $h=\delta \omega_{\text{max}}/ \omega_0 \times \tau$ by
\begin{equation}
\label{eq:h-thinwall}
h = A^2\Gamma_{\text{eff}} \tau \, ,
\end{equation}
 $\tau = d/v_\perp $ being the interaction time between the wall moving at velocity $v$ and an individual device. Again, for the wall to be ``thin'', we require $\tau$ to be less than the sampling time interval $\tau_0$.

With the theoretical background established, we now review GPS data and characterize the utility of applying the  matched-filter technique for DM search in network data streams. This includes establishing a signal-to-noise ratio test statistic and benchmarking the method via simulation.

\section{Overview of GPS data}\label{sec:data}

A detailed description of modern GPS data acquisition and processing techniques and their application in precision geodesy can be found in Ref.~\cite{treatise}. Details relevant to DM searches with GPS constellation
are given in Ref.~\cite{Roberts2017}.
Here, we briefly review the main aspects of GPS atomic clock data and introduce relevant concepts and terminology.

In our search, we analyze the GPS data generated by the Jet Propulsion Laboratory (JPL)~\cite{JPL}. 
This data consists of clock biases, the difference in clock phases (i.e., the operational ``time'' as counted by the clocks) between a given satellite clock and a fixed reference clock, and are sampled at $\tau_0= 30 \units{s}$ intervals. 
The same reference clock is used for the entire network of satellite clocks on any given day.
The data set also provides the satellite orbits, so we know the locations of the networks nodes (satellites).  
The JPL performs the initial GPS data processing~\cite{JPL}. In their processing, they do not limit clock bias behavior, meaning that real transient signals are not removed as outliers.

Each clock's bias data, denoted $d_j^{(0)}$, is non-stationary and is dominated by random walk process. 
Prior to our analysis, we ``whitten'' the data by performing the first-order differencing and define a new data stream
\begin{equation}
    d_j^{(1)}=d_j^{(0)}-d_{j-1}^{(0)}.
\end{equation}
This differencing procedure is sufficient for the Rb satellite clocks while a second-order differencing procedure ($d_j^{(2)}=d_j^{(0)}-2d_{j-1}^{(0)}+d_{j-2}^{(0)}$) is often preferred for Cs clocks.
We refer to the differenced data $d_j^{(1)}$ as the pseudo-frequency due to its proportionality to the discrete clock bias derivative. The units of pseudo-frequency $d_j^{(1)}$ are nanoseconds. As shown in Ref.~\cite{Roberts2018b} the pseudo-frequency noise is dominated by
the Gaussian white noise.
To streamline notation, for the rest of the paper, $d_j \equiv d_j^{(1)} $. Such data standard deviation $\sigma$ is related to the commonly used Allan deviation $\sigma_y(\tau_0)$ as  $\sigma = \tau_0 \sigma_y(\tau_0)$. 

An important aspect of the GPS time series data is that it consists of the individual clock noise and the noise of the reference clock.
So, for each clock $a$, the noise component can be represented as
\begin{equation}\label{eq:timeseries}
    n_j^a = e_j^a - c_j \, ,
\end{equation}
where $e_j$ is the individual clock noise and $c_j$ is the contribution from the reference clock noise common to all data streams. Here and below the subscript enumerates data points (epochs) and the superscript enumerates sensors. While both sources of noise in pseudo-frequencies are dominated by the Gaussian white noise, in our simulations we will include realistic auto-correlation functions for the GPS clocks computed in~\cite{Roberts2018b}.

\subsection{Simulating GPS data}
Characterizing the efficacy of the MFT is contingent on our ability to simulate the GPS atomic clock data. 
A detailed description of GPS data simulation along with a direct comparison to the GPS archival data is provided in Ref.~\cite{Roberts2018b}.
The essence of the simulation method comes from utilizing the known power spectral densities for each clock (from JPL) to ``color'' pseudo-random white noise~\cite{Rollings}. 
Moreover, we are able to simulate cross-clock correlation by adding an extra set of simulated white noise with standard deviation equal to that of a typical reference clock ($\sigma_\times \approx 0.006~\un{ns}$) to all of the simulated satellite clock data streams.
This effectively acts as the common reference clock noise component to the satellite data in Eq.~(\ref{eq:timeseries}).

With the necessary background provided, we pivot to a description of our methodology in next section.

\section{Methodology}\label{sec:method}
We wish to determine whether there is significant evidence that a thin wall DM signal is present in the GPS archival atomic clock data or not.
This can be formalized by the following two-sided hypothesis test:
$$
H_0: h = 0, \quad \mathrm{vs.} \quad H_1: h \neq 0 \, ,
$$
where $h$ represents the strength of the possible hidden DM signal. 
Note that the alternative hypothesis $H_1$ can be thought of a union of all possible alternatives $H_h$ where $h$ may be any non-zero real number. 
Furthermore, note that we allow the signal strength $h$ to be both positive and negative, which is different than a typical signal strength search that treats $h$ as an amplitude (and therefore non-negative).
This is because the sign of the DM interaction coupling $\Gamma$ is not known {\em a priori}. That is, we do not know if the DM-induced perturbations will cause the GPS atomic clocks to tick faster or slower.

To perform the aforementioned hypothesis test, we must formulate a detection statistic $\rho$, which in this case will be a signal-to-noise ratio and is formulated in the following section. Then, in order to claim detection, we must first establish a detection threshold $\rho^*$ (provided in Sec~\ref{sec:thresh}) to compare to our observed statistic $\rho$. 

The cases in which our search produces an observed SNR that exceeds the established threshold or not are treated differently. 
If our search results in a detection statistic larger than the threshold, we then wish to estimate the parameters relating to the DM interaction event. 
This is discussed in Sec~\ref{sec:param}. 
On the other hand, if our search does not result in a detection statistic indicating an event, we wish to place limits on the signal strength $h$ which translates into into limits on the DM coupling $\Gamma$ via Eq.~(\ref{eq:h-thinwall}), see  Sec~\ref{sec:reach}.

\subsection{Formulation of test statistic}\label{sec:snr}
Consider a candidate DM model $M$ that would leave a coherent signal in network sensor data set $D$.
The data set may or may not include a candidate model-prescribed signal $\bm{s}(\bm{\theta})$, where $\bm{\theta}$ is the specific set of parameters that define the signal signature, such as the DM object's velocity, orientation, arrival time and strength. If there is no signal within the data, then $\bm{s}=0$.
Taking a frequentist approach, all one needs is a likelihood function for the data set. This is given by the following Gaussian:
\begin{equation}\label{eq:like}
    \mathcal{L}(D|M) = K \exp \Big[ -\frac{1}{2} \chi^2(\bm{s}) \Big] \, ,
\end{equation}
where $K$ is the normalization factor and
\begin{equation}
     \chi^2(\bm{s})=\sum_{ab}^{\ND} \sum_{jl}^{J_W} \big( d_j^a-s_j^a\big) \big(E^{-1}\big)_{jl}^{ab} \big( d_l^b-s_l^b\big) \, .
\end{equation}
Here $d_j^a$ is data for the $a^{th}$ clock at epoch $j$ and $\bm{E}$ is the covariance matrix for the network.
To streamline notation, we dropped the explicit reference to template parameters $\bm{\theta}$. The indices $a,b$ run over $\ND$ sensors (excluding the reference clock for GPS)  and $j,l$ range over the epochs (data points) in the observation time window of length $J_W$. In an equivalent vector notation, 
\begin{equation}
 \chi^2(\bm{s})=(\bm{d}-\bm{s})^T\bm{E}^{-1}(\bm{d}-\bm{s}) \, ,
\end{equation}
where $\bm{d}$ is the data stream and $\bm{s}$ is the signal stream.

The likelihood in Eq.~(\ref{eq:like}) is a multivariate function of the signal parameters $\bm{\theta}$. An important aspect of our method is the signal linearity  with respect to its  strength $h$ so that we can define a ``unit'' signal that is scaled by its strength
\begin{equation}
    \bm{s}(\bm{\theta}) = h\bar{\bm{s}}(\bm{\theta}|_{h=1}) \, .
\end{equation}
This way, our likelihood becomes
\begin{equation}\label{eq:like_w_h}
    \mathcal{L}(D|M) = K \exp \Big[ -\frac{1}{2} \Big( (\bm{d}-h\bar{\bm{s}})^T\bm{E}^{-1}(\bm{d}-h\bar{\bm{s}}) \Big) \Big] \, .
\end{equation}

Since we do not know the shape or parameters of the unit signal $\bar{\bm{s}}$ (or if there exists a signal within the data at all), we span the space of all possible signals by forming a large repository of unit signal templates (a discussion of how we form the repository of templates is provided in Sec.~\ref{sec:mc}). Suppose we form a unit signal template $\bar{\bm{s}}_{i}$ with a set of randomly (but strategically) chosen parameters. We then compare this template to the data stream via the likelihood function in Eq.~(\ref{eq:like_w_h}). The template specific likelihood is then a function of only the signal strength $h$ since each of the other signal parameters have been fixed to form $\bar{\bm{s}}_{i}$. In this case, the template-specific likelihood can be re-cast as a function of $h$ alone
\begin{equation}
    \mathcal{L}_i(D|M) \propto \exp \Big[ -\frac{1}{2} \Big( \frac{h-\hat{h}}{\sigma_{h}} \Big)^2 \Big] \, ,
\end{equation}
where
\begin{align}
    \hat{h} &= \frac{\bm{d}^T\bm{E}^{-1}\bar{\bm{s}}_i}{\bar{\bm{s}}_i^T\bm{E}^{-1}\bar{\bm{s}}_i} \, ,\\
    \sigma_{h} &= \frac{1}{\sqrt{\bar{\bm{s}}_i^T\bm{E}^{-1}\bar{\bm{s}}_i}} \, .
\end{align}
Here $\hat{h}$ is the signal strength that maximizes the template-specific likelihood and $\sigma_h$ is the template-specific likelihood standard deviation.
We quantify how well the signal template $\bar{\bm{s}}_{i}$ matches the signal within the data via a signal-to-noise ratio statistic defined as
\begin{equation}\label{eq:gensnr}
    \rho_i \equiv \frac{\hat{h}}{\sigma_{h}} = \frac{\bm{d}^T\bm{E}^{-1}\bar{\bm{s}}_i}{\sqrt{\bar{\bm{s}}_i^T\bm{E}^{-1}\bar{\bm{s}}_i}} \, .
\end{equation}

We  emphasize that our  use of the SNR as a statistic rather than its square is to retain the dependence on the sign of the DM coupling. The SNR statistic depends on the inverse of the covariance matrix $\bm{E}$. Properties of the covariance matrix and its inversion techniques are discussed in Appendix~\ref{app:covarInv}.
Note that Eq.~(\ref{eq:gensnr}) is general in that it applies to any modelled signal (monopoles, strings, walls, etc.), though here we treat the case of thin walls only.

In general, due to the central limit theorem, we can assume that the intrinsic noise of the network sensors is Gaussian (the noise may be colored). This was the underlying assumption in writing the likelihood (\ref{eq:like}).  Then the SNR statistic~(\ref{eq:gensnr}) is a linear combination of Gaussian random variables, and as such, the SNR is also a random Gaussian variable.  

Now, recall that we wish to span the space of possible model-prescribed signals with a repository of unit signal templates. Then, given a repository of $M$ randomly generated templates, we define our detection statistic $\rho$ as the template-specific SNR with the largest magnitude
\begin{equation}\label{eq:detectstat}
    \rho = \rho_j \quad \textrm{such that} \quad |\rho_j| =\max\{ |\rho_i| \}_{i=1}^M \, .
\end{equation}
Choosing this as the detection statistic results in finding the signal template that
maximizes the multivariate likelihood~(\ref{eq:like}). 

With our detection statistic defined, in the next sections we outline our method of unit template generation and provide an overview of our procedure for searching the GPS datastreams for DM events.

\subsection{Template generation}\label{sec:mc}

Each signal template is determined by the DM model used (in this case, the thin domain wall) and the necessary parameters associated with the event: velocity (and incident direction), time of the event and thickness of the DM object. 
Since it would take an infinite number of model-prescribed signal templates to cover such a continuous parameter space, we strategically generate our finite repositories (template banks) of signals with a Monte-Carlo approach using prior distributions for individual parameters.
When generating signals for a template bank, we employ importance sampling for each parameter according to these prior distributions in an effort to approximately span the continuous parameter space with a finite sample.
This approach is formalized in appendix~\ref{app:ITS}.

We use the SHM to generate necessary parameter prior distributions. 
The velocity distribution for DM objects in the halo is quasi-Maxwellian and isotropic with a dispersion around $v\approx300 \, \textrm{km s}^{-1}$~\cite{BOVY2012}. 
In addition, there is a motion of the Solar system through the halo at galactic velocities of $v_g \approx 220 \, \textrm{km s}^{-1}$.
The resulting most probable incident direction of a DM object is along the path of the Sun's orbit in the galaxy, toward the Cygnus constellation. This implies that over 90\% of DM events would come from the forward facing hemisphere~\cite{Roberts2017}. 
Further discussion of priors for event parameters such as domain wall width and event rate is provided in Ref.~\cite{Roberts2018b}. 
The event parameters (velocity, incident direction, etc.) determine in which order the nodes are swept, as well as how quick the sweep is. These characteristics distinguish the templates within the template bank.

\begin{figure}[t!]
    \centering
    \includegraphics[width=\linewidth]{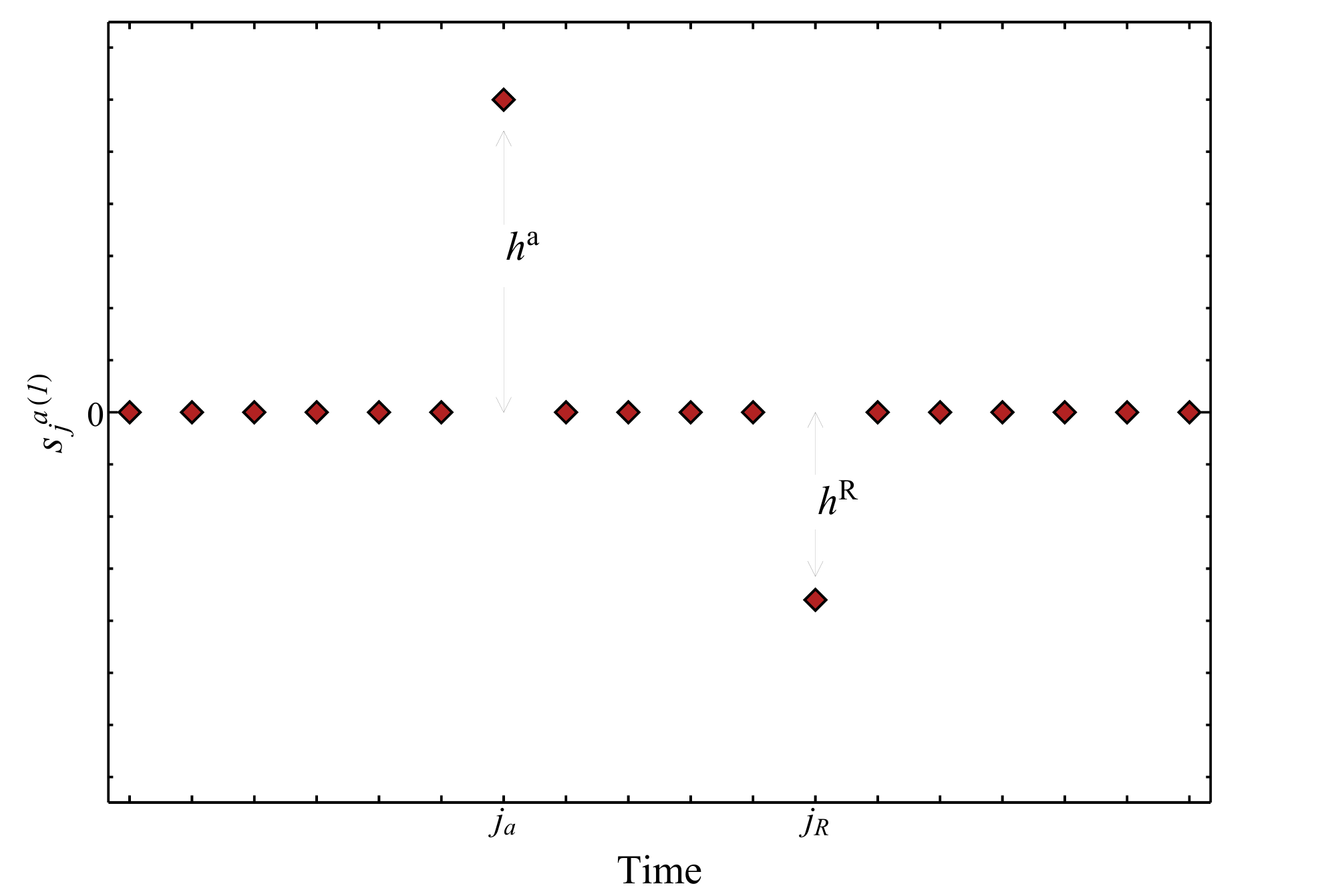}
    \caption{Time series for a differenced thin domain wall signal. Here clock $a$ is affected by the DM object prior to the reference clock $R$. When the thin wall object interacts with clock $a$ between epochs $j_a-1$ and $j_a$, a spike of magnitude $h^a$ is seen at epoch ${j_a}$. Then, as the thin wall object sweeps the reference clock  between epochs ${j}_{R}-1$ and $j_R$, a spike of opposite sign and magnitude $h^R$ is seen at epoch $j_R$.}
    \label{fig:signal}
\end{figure}

After generating the template parameters, we form the signal templates. The thin domain wall is characterized by an interaction time with each device (clock) of less than the sampling time interval; thereby, the profile  contains delta-functions of time.
The DM-induced clock bias (phase) of a given clock $a$ is proportional to an integral of the frequency shift~(\ref{eq:variation}). 
Further, the bias data stream is given with respect to a fixed reference clock $R$, which is also affected by the domain wall. We thus distinguish between maximum signals $h^a$ and $h^R$ as shown in Eq.~(\ref{eq:h-thinwall}). 
The signal in the differenced data stream ${s^a_j}^{(1)}$ then reads (for the case when the wall encounters the clock $a$ prior to the reference clock $R$)
\begin{equation}
	{s^a_j}^{(1)}=
	\begin{cases}
		0 ,		& j < j_a \,,  \\
		h^a , 	& j = j_a \,,  \\
		0 ,     & j_a < j < j_R \,,  \\ 
		-h^R ,	& j = j_R\,,   \\
		0,	    & j > j_R\,,  \\
	\end{cases}
\label{eq:sij-thinwall}
\end{equation}
where the time at epoch $j$ is $j\times \t_0$, a discrete time on the sampling grid, and $j_a,j_R$ are the epochs in which the satellite clock and reference clock interact with the DM object, respectively.
This template is shown graphically in Fig.~\ref{fig:signal}.

In a homogeneous network of clocks, the values of $h^a$ and  $h^R$ are the same, thereby allowing us to split a single $h$ from the differenced signal to form the templates with unit spikes at $j_a$ and $j_R$.
This is also true for any network under the assumption that the $\Gamma_{m_e}$ coupling dominates over other couplings (i.e., $|\Gamma_{m_e}| \gg |\Gamma_{\a,q}|$) since the $\Gamma_{m_e}$ contribution in Eqs. (\ref{eq:Crb} - \ref{eq:Ch}) is the same for all clock types. 
Assuming that any other coupling dominates, we may still split a single $h$ from the differenced signal but the unit templates will contain a unit spike at $j_a$ and a spike of magnitude $\eta \equiv h^R/h^a = \Gamma_{\text{eff}}^R/\Gamma_{\text{eff}}^a$ and of opposite sign at $j_R$.

We would also like to highlight the importance of a well-spaced network for the MFT approach and template generation. Well-spaced meaning that few (if any) satellite clocks are affected by the DM wall within the same $30\un{s}$ time period as the reference device. If the network nodes were not sufficiently spatially separated, the signal templates from Eq.~(\ref{eq:sij-thinwall}) would collapse into ``null'' templates, where all elements of the signal stream are zero. This is due to the node devices and reference device de-synchronizing and re-synchronizing all within the time period of one epoch, effectively eliminating detectable DM interaction effects on the data stream. 

Beyond individual template generation, we must choose an appropriate number of templates in the repository to accurately span the parameter space. In Sec.~\ref{sec:dp} we gauge how the number of templates affects the DM signal detection capabilities.

\section{Analytic results for idealized network}\label{sec:TWsnr}

Now we turn to the general SNR~(\ref{eq:gensnr}) and determine the statistical properties of SNR for thin domain wall signals~(\ref{eq:sij-thinwall}). 
In this section, we consider an analytically treatable case of an idealized network comprised of $\ND$ identical white noise sensors.  We additionally incorporate a white noise reference sensor common to all the sensors. This common noise reference sensor is especially relevant to GPS clock network, where it arises due to all clock biases reported with respect to a single reference clock. We will denote the intrinsic noise variance of the network sensors and the reference sensor  as $\sigma^2$ and $\sigma_{\times}^2$, respectively. Both the sensors and the reference  can be affected by dark matter transients.

We will determine the expected distribution of the template-specific SNR values $\rho_i$ given that there is no signal in the data stream, $\Prob{(\rho_i|H_0)}$, as well as the distribution of the detection statistic given that there is a signal of strength $h$ present, $\Prob{(\rho|H_h)}$, for this idealized sensor network.

As discussed in Sec.~\ref{sec:mft}, the central quantity of interest, SNR~(\ref{eq:gensnr}), is a Gaussian random variable and as such its probability distribution is fully characterized by its mean value and variance. Because it is random, the SNR can fluctuate. Due to these fluctuation, even in the absence of the DM signal, the SNR may attain large values that can be falsely misinterpreted as the presence of the DM signal. The larger the SNR variance, the larger the fluctuations are, and the larger detection threshold must be.

For the idealized network, the inverse of the covariance matrix needed to compute the SNR statistic can be found analytically (see Appendix~\ref{app:covarInv})
\begin{equation}
    \Big( E^{-1} \Big)^{ab}_{jl} = \frac{1}{\sigma^2}\delta_{jl}\Big(\delta^{ab}-\frac{1}{\ND} \frac{\xi}{1+\xi} \Big) \,,
\end{equation}
where $\xi \equiv \ND\sigma_{\times}^2/\sigma^2$. 

Now, if there is a signal present in the data stream, each individual sensor's data is given by $d_j^a = e_j^a - c_j + h\bar{s}_j^a $ (a sum of an individual sensor noise, reference  noise and a signal term). When the signal is absent, one can simply set $h\to 0$. 
Our explicit computation using Eq.~(\ref{eq:gensnr}) with $\bar{\bm{s}}_i = \bar{\bm{s}}$  (see Appendix~\ref{app:twsnr-deriv} for derivation) results in a Gaussian distribution for $\rho$ with a mean of 
\begin{equation}\label{eq:twmean}
   \mu_{\rho} = \frac{h\sqrt{\ND}}{\sigma}\sqrt{\frac{1+\eta^2+\xi}{1+\xi}} \,
\end{equation}
and variance of
\begin{equation}\label{eq:twvar}
     \sigma_{\rho}^2 = 1 \, .
\end{equation}
Here $\eta \equiv h^R/h^a = \Gamma_\mathrm{eff}^R/\Gamma_\mathrm{eff}^a$ is the ratio between the strength of the signal experienced for the reference clock to that of the satellite clocks. 
Note that we assumed that device degeneracy (multiple sensors experiencing a signal at the same epoch) can be ignored. 
Note that when the DM signal is absent ($h=0$), $\mu_{\rho} =0$ while $\sigma_{\rho}$ remains constant.
Moreover, the standard deviation of template-specific SNR $\sigma_{\rho_i}$ is also constant at 1. 
The probability density distribution for SNR (for a fixed, matching template) is given by
\begin{equation}
    \Prob(\rho|H_h ) = \frac{1}{\sqrt{2\pi} }
    \exp{ \left[-\frac{(\rho -\mu_\rho )^2}{2 } \right] } \, .
    \label{Eq:SNR-prob-distro}
\end{equation}

Assuming that none of the couplings $\Gamma_X$ dominates the DM interaction with GPS devices, $\eta \approx 1$ for any of the satellite-reference clock combination (see Eqs.~(\ref{eq:Crb}-\ref{eq:Ch})). This remains true if either $\Gamma_\alpha$ or $\Gamma_{m_e}$ dominates the interaction. The only major deviation of $\eta$ from a value near $1$ is when $\Gamma_{m_q}$ is the dominate coupling, for which the use of a network of Rb clocks with an H-maser reference clock will result in $\eta \approx 2$. For the following analysis, we will assume that $\eta = 1$.

In the limit  $\xi \ll 1$, i.e., $\sigma_\times \ll \sigma/\sqrt{\ND}$, we arrive at $\mu_{\rho} = h\sqrt{2\ND}/\sigma$, recovering the known result for a network of uncorrelated devices  (see e.g., Ref~\cite{RomanoCornish2017}). For networks with large cross-correlation or large number of devices ($\xi \gg 1$), we arrive at $\mu_\rho =  h\sqrt{\ND}/\sigma$, a factor of $\sqrt{2}$ less than the uncorrelated network. Regardless of the level of cross-correlation, the network sensitivity grows with the sensor number as $\sqrt{\ND}$.

In our search, we do not use the exact inversion of the covariance matrix as was used to derive the expressions in this section. Instead we incorporate a perturbative inversion (see Appendix~\ref{app:covarInv}) which assumes that the reference clock noise is small compared to the noise of the satellite clocks. 

\subsection{Multiple events}
The Bayesian technique outlined in \cite{Roberts2018b} assumed there to be at most one DM interaction event in any particular time window of the archival GPS data. So far, here we have also only treated the case of a single thin wall interaction event occurring in a time period of $J_W$ epochs. However, if we consider dark matter encounters to be Poisson distributed in time, with an average time between consecutive events $\CT$, over the 20 years of archival data we would expect there to be $N_E = ( 20 \,  \mathrm{years})/\CT$ events. Then, extending our search window $J_W$ to contain the total number of epochs in the entire two-decade window of GPS data, and assuming that consecutive events are non-overlapping, we find that the mean of our detection statistic (\ref{eq:twmean}) increases by a factor of $\sqrt{N_E}$, while the variance of the statistic remains unchanged. This ultimately improves our sensitivity by $\sqrt{N_E}$.

While this section analyzed an idealized network, simplifying assumptions will be lifted in full numerical simulations in later sections. We will use real colored noise auto-correlation functions for heterogeneous networks of GPS clocks.


\section{Determining a detection threshold}\label{sec:thresh}
In order to find a detection threshold, we must determine how the detection statistic behaves in the case when there is no signal present in the data.
This way we can determine whether the computed statistic provides a significant evidence for rejecting the null hypothesis and claiming  DM detection.
Rather than obtaining the distribution for our test statistic given the null hypothesis is true, $\Prob(\rho|H_0)$, we determine our detection threshold in a different yet equivalent fashion given the nature of our test statistic.

\begin{figure*}[ht!]
    \centering
    \includegraphics[width=\textwidth]{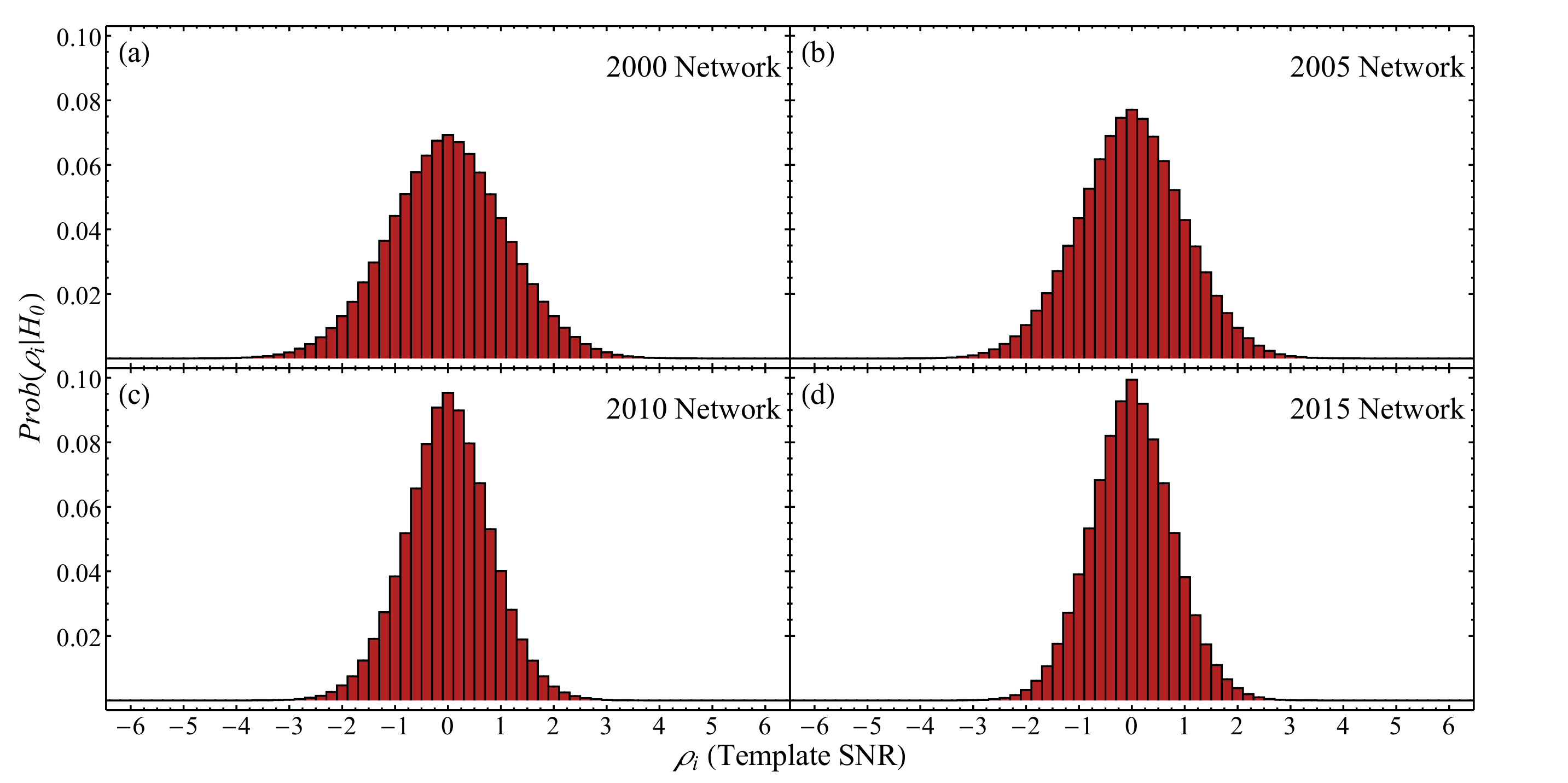}
    \caption{Histograms of template specific SNR calculations ($\rho_i$) on signal-free simulated data for various clock networks. The network composition for indicated years is given in Table~\protect\ref{tab:table1}. 
Standard deviations for template-specific SNR values are: 1.15 (2000), 1.03 (2005), 0.86 (2010), and 0.81 (2015). }
    \label{fig:noSignal}
\end{figure*}

Recall that we define our detection statistic as the template-specific SNR with the \textit{maximum magnitude} out of a repository of $M$ templates. Instead of assessing the distribution of the maximum magnitude $\rho_i$, we simply assess the distribution of $\rho_i$. To this end, we performed Monte-Carlo simulations consisting of $\approx10^6$ SNR calculations on event free simulated data and confirmed that the distributions for $\rho_i$ given the null hypothesis is true are Gaussian with a mean of zero and standard deviation $\sigma_{\rho_i}$. 
The results of the simulations for various simulated clock networks (clock networks for the years 2000, 2005, 2010, and 2015; see Table~\ref{tab:table1}) are provided in Fig.~\ref{fig:noSignal}.
Given that the template-specific SNR behaves in this fashion, the probability that any of the templates in the repository produces an SNR value larger in magnitude than some SNR threshold $\rho^* = n^*\sigma_{\rho_i}$, is
\begin{equation}\label{eq:nstar}
\Prob(|\rho_i|>n^*\sigma_{\rho_i}|H_0)= 1-\Big[ \text{erf} \Big(\frac{n^*}{\sqrt{2}}\Big) \Big]^M \, ,
\end{equation}
where $M$ is the number of templates used in the template repository. This is a false positive rate per epoch. 

\begin{table}[ht!]
\caption{\label{tab:table1} GPS network clock types used for simulation with a white noise reference clock.}
\begin{ruledtabular}
\begin{tabular}{ccccccccc}
 Year & Cs-II & Cs-IIA & Cs-IIF & Rb-II & Rb-IIA & Rb-IIR & Rb-IIF & H\footnotemark[1]\\ \hline
 2000 & 5 & 11 & 0 & 1 & 7 & 3 & 0 & 0\\
 2005 & 1 & 8 & 0 & 1 & 8 & 12 & 0 & 0\\
 2010
 & 0 & 5 & 0 & 0 & 5 & 19 & 2 & 2\\
 2015
 & 0 & 1 & 1 & 0 & 3 & 19 & 3 & 0\\
\end{tabular}
\end{ruledtabular}
\footnotetext[1]{These are terrestrial H-maser clocks.}
\end{table}

A reliable SNR threshold will ensure that we can expect less that 1 false positive in $Z$ epochs. The value of $n^*$ that meets this criterion is
\begin{equation}
n^* = \sqrt{2} \, \text{erf}^{-1} \Big[ \Big( 1-\frac{1}{Z} \Big)^{\frac{1}{M}} \Big] \approx \sqrt{\text{Log}
\Big(\frac{2M^2Z^2}{\pi}\Big)} \, . 
\end{equation}

The most reliable threshold would allow less than one false positive in the entire span of data. For 20 years of archival 30-second GPS data, $Z=2.1\times10^7$ epochs. With $M=1024$ templates in the repository, the value of $n^*$ is 6.57, corresponding to a threshold SNR of $\rho^* = 6.57\sigma_{\rho_i}$. 
However, less strict detection thresholds may be used to identify possible weak candidate events for further investigation.
Note that the value for $n^*$ depends weakly (logarithmically) on the number of templates $M$, thereby it   does not vary significantly for different sized template repositories. 
Ultimately, our detection threshold is given by $\rho^* = n^* \sigma_{\rho_i}$. 
Using the distributions in Fig.~\ref{fig:noSignal}, we calculate the detection thresholds for each network allowing for 1 false positive per day, 10 false positives per year, and 1 false positive in 20 years. Table~\ref{tab:table2} summarizes the results.

\begin{table}[ht!]
\caption{\label{tab:table2} Detection thresholds $\rho^*$ as a function of various generations of GPS constellation and  false positive (f.p.) rates. The network compositions for indicated years are given in Table~\protect\ref{tab:table1}. 
}
\begin{ruledtabular}
\begin{tabular}{cccc}
 Year & 1 f.p./day &  10 f.p./year & 
 1 f.p./20 years \\
\hline
2000 & 5.87 & 6.61 & 7.58 \\ 
2005 & 5.26 & 5.92 & 6.79 \\
2010 & 4.40 & 4.95 & 5.67 \\
2015 & 4.11 & 4.62 & 5.30 \\
\end{tabular}
\end{ruledtabular}
\end{table}

\subsection{Future networks and alternative data processing}\label{sec:mitigation}

As mentioned in Sec.~\ref{sec:TWsnr}, in our simulations we use a perturbative inverse of the covariance matrix; see appendix~\ref{app:covarInv}. 
This approximation relies on the noise level of the satellite clocks being sufficiently larger than the noise of the reference clock.
The GPS clock networks for the years 2000 through 2015 satisfy the quiet reference clock requirement.
In recent years, however, more stable Rb-IIF clocks with noise comparable to that of the reference clocks have been added to the GPS constellation thereby weakening the justification for the perturbative approximation for $\bm{E}^{-1}$.
Moreover, future GNSS networks will contain a plethora of ground- and satellite- based H-maser clocks to be exploited in our searches (Galileo satellites already house stable H-masers~\cite{GNSS}).
Switching our method to use the exact inversion mitigates, but with the trade-off of computational overhead.
We wish to avoid using an exact inversion of the covariance matrix due to the fact that it drastically increases computation time and would add a considerable amount of time to a search through the GPS data.

Considering the insufficiency of the perturbative approximation for more recent clock networks, here we offer a possible mitigation technique.
As more accurate satellite clocks are being placed in orbit, more reference clocks are being placed around the globe. 
We propose pairing each of the satellite clocks with their own reference clock, thereby eliminating the cross-correlation caused by the use of a single reference clock that is inhibiting current search techniques. 
Suppose there are $N_D$ atomic clocks, satellite- and Earth-based, at our disposal with half of them being Earth-based. 
The large level of cross-correlation that restricts the perturbative inversion may be eliminated by using data from $N_D/2$ satellite-Earth clock pairs.
The application of the matched-filter technique can be reformulated for a network of device pairs and is left for future work when such networks become a reality. 

\section{Detection Probability}\label{sec:dp}
\subsection{Detecting events}
In the event of a weak DM signal presence in the data stream, it may not be immediately noticeable in the atomic clock data due to the randomness of the clock noise. 
The advantage of the SNR (and a detection statistic in general) is to provide a clear gauge of the signal presence.
We verify that the SNR statistic is capable of detecting sought DM signals via simulation. 

\begin{figure}[ht!]
    \centering
    \includegraphics[width=\linewidth]{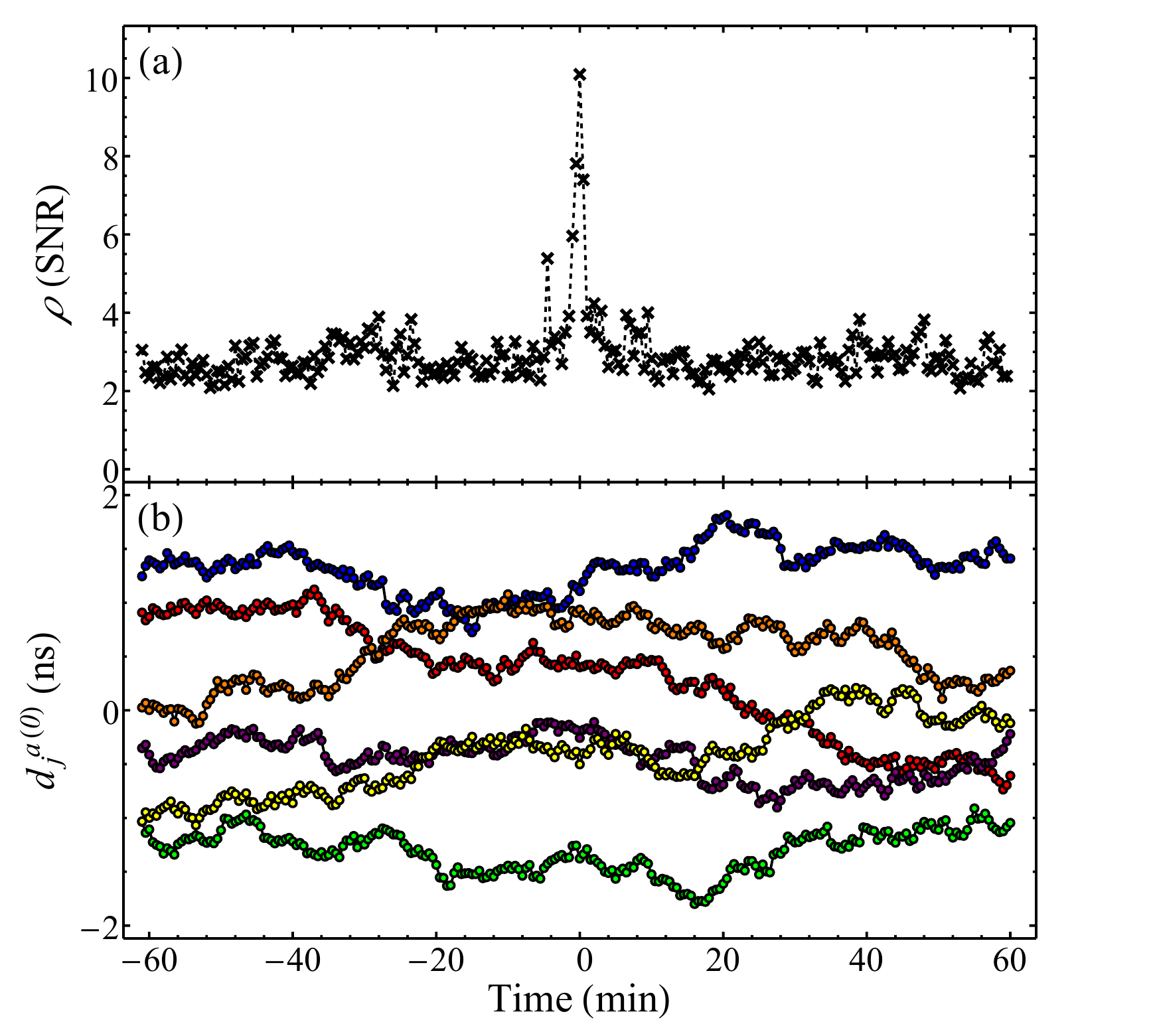}
    \caption{Frequentist detection of a simulated thin domain wall signal. The thin wall with strength $h=0.10 \un{ns}$ interacts with a network of 30 white noise clocks, each with $\sigma=0.05 \un{ns}$, at time $t_0=0$. The common reference clock contribution is white noise with $\sigma_\times = 0.006~\un{ns}$. {\em Bottom panel}: simulated clock bias data for 6 of the 30 clocks with the injected signal included. For visual clarity, each data stream is shifted by a constant. {\em Top panel}: magnitude of the corresponding SNR statistic $\rho$ for the same time scale.}
    \label{fig:singleStream}
\end{figure}

To this end, we simulated 2 hours of data for a network of $N_D=30$ clocks that exhibit Gaussian white noise with a standard deviation of $\sigma = 0.05~\un{ns}$ along with a white noise reference clock contribution with noise level $\sigma_\times = 0.006~\un{ns}$.
We then injected a signal of strength $h=0.1\,\un{ns}$ in the middle of the data stream with normal velocity of $v_\perp\approx300 \un{km}\un{s}^{-1}$ and incident direction angles $\theta \approx 1.7\pi \un{rad}$ and $\phi \approx 0.2\pi \un{rad}$ (this in the Earth-centered Inertial (ECI) J2000 frame), which are the most probable event parameters according to the SHM and our previous calculations (see \cite{Roberts2018b}).
The results of performing our search technique on this data set is shown in Fig.~\ref{fig:singleStream}, where we show the time series data (with the injected signal) for 6 of the 30 clocks. 
The upper panel shows magnitude of the calculated detection statistic [Eq.~(\ref{eq:detectstat})] for each epoch in the two hour window.
This simulation used a template repository of size $M=1024$ and the calculated detection threshold for this type of network is given by $\rho^* = 6.57$ (allowing for no false-positives in 20 years).
While the injected signal is not recognizable by the eye in the simulated data streams, a spike in the detection statistic at the time of the injected event is apparent.

Note that the search method is not aware of the event's strength, speed, direction, time of occurrence, or the fact that there was an injected signal at all.  
The injected signals were generated independently of the search routine and the template bank. 

\subsection{Detection probability}
The main figure of merit of the MFT algorithm is a detection probability curve for the various clock networks that have been collecting data for the past two decades. The detection probability is defined as the probability that the observed detection statistic exceeds the detection threshold given that the alternative hypothesis is true: $\Prob(\rho > \rho^* |H_h)$. 
We wish to determine the detection probability for our various clock networks as a function of the signal strength $h$ and  obtain a $95\%$ detection probability signal strength, denoted $h^{95\%,\,\text{D.P.}}$.

\begin{figure}[t!]
    \centering
    \includegraphics[width=\linewidth]{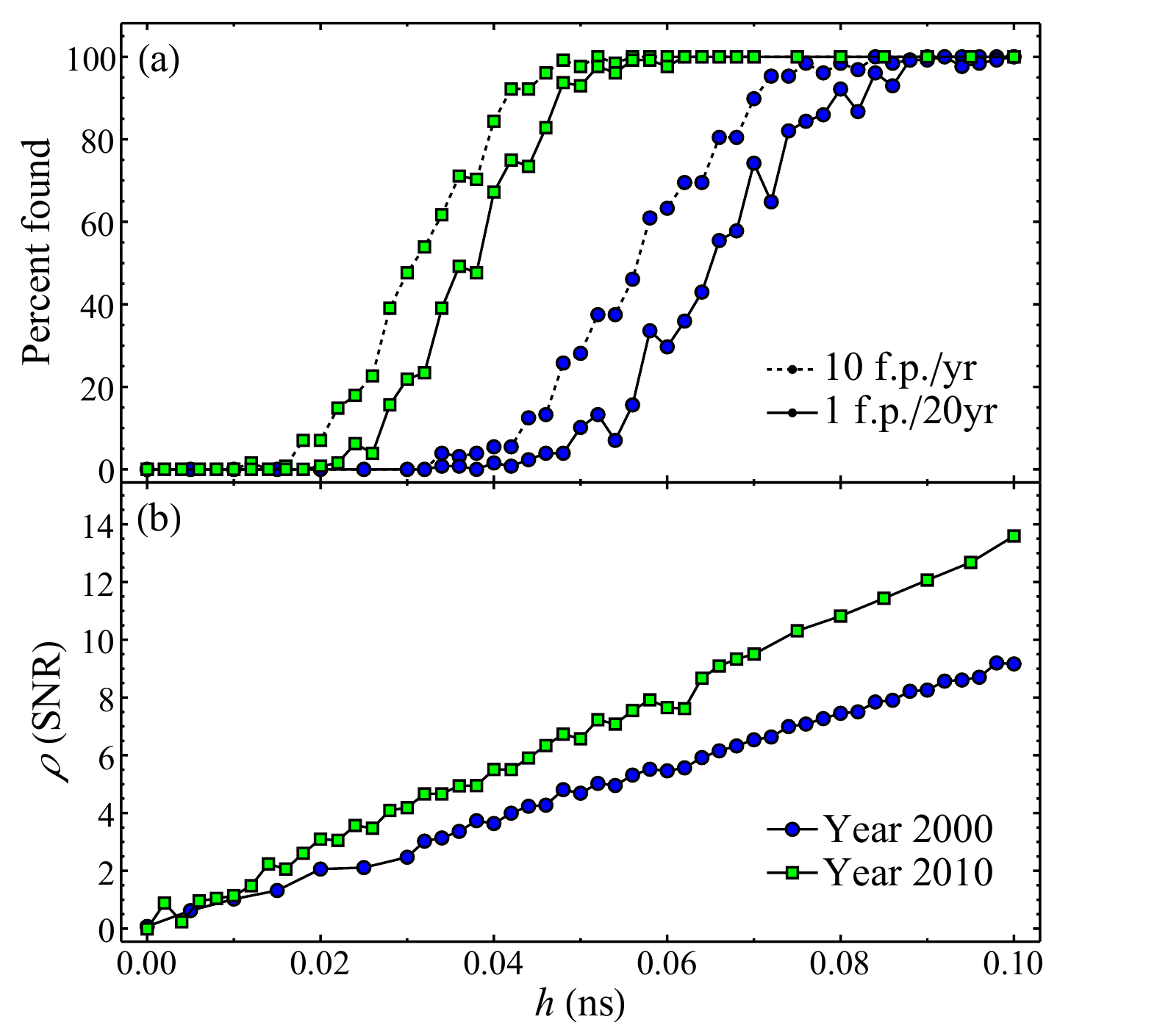}
    \caption{\emph{Top panel:} Detection probability curves as a function of the injected signal strength $h$ for various simulated clock networks. Each point represents the percent of injected events of strength $h$ found for 128 trials. Each curve contains 42 points. \emph{Bottom panel:} The average detection statistic value $\rho$ of the 128 trials at the time the event was injected. Though not plotted here, the 2005 network maintains a similar sensitivity to the 2000 clock network while that of the 2015 clock network sensitivity is similar to the 2010 network.}
    \label{fig:percFound}
\end{figure}

Monte-Carlo simulations produced the detection probability curves shown in Fig.~\ref{fig:percFound}. 
The simulation scheme consisted of $128$ trials where a randomly-generated thin-wall signal of strength $h$ was injected into a data stream for a particular clock network and the detection statistic was calculated for every epoch within the simulated data stream. 
An event was considered found if the computed SNR  exceeded the network's threshold within one epoch of the injected event time.
We compared the calculated SNR values with two  different detection thresholds: one that allows for 10 false positive events per year and another one that allows for less than 1 false positive event in the 20 year span of the GPS data.
The number of found events divided by the number of iterations gave us the detection probability. 
This detection probability was computed for a range of injected signal strengths.

Along with the detection probability curves, we plot the average of the $128$ SNR calculations at the epoch where the signal was injected as a function of the signal strength. 
This is also shown in Fig.~\ref{fig:percFound}. We can see that the SNR is a linear function of $h$, as expected. 
This fact helps form a confidence interval for the signal strength $h$ in the event that a DM signal is found.

\begin{figure}[ht!]
    \centering
    \includegraphics[width=\linewidth]{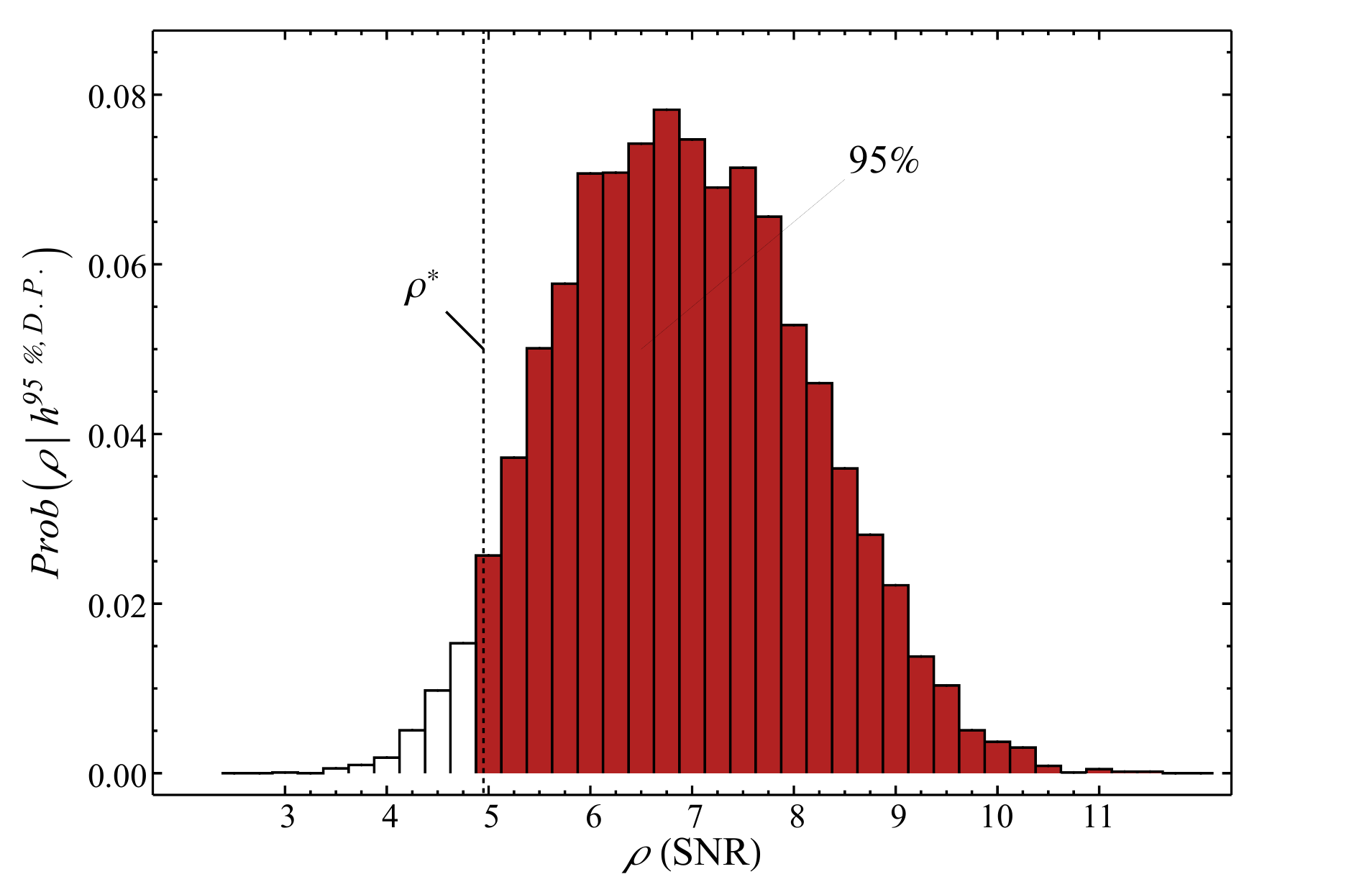}
    \caption{Histogram of the detection statistic for a simulated network replicating the clock types of the 2010 GPS constellation with an injected signal of strength equal $h^{95\%,\,\text{D.P.}}$ ($h=0.045\un{ns}$). The dotted black line is the detection threshold for this network ($\rho^* = 4.95$). The shaded red region encompasses all iterations that resulted in a detection statistic larger than the threshold. $94.6\%$ of the iterations fall into this region. }
    \label{fig:yesSignal}
\end{figure}

To verify that our detection probability curve provides us with the correct value for $h^{95\%,\,\text{D.P.}}$, we performed an auxiliary simulation. 
Here we injected signals of strength $h^{95\%,\,\text{D.P.}} = 0.045\un{ns}$ for the 2010 clock network into a simulated data stream. For each signal injection with random parameters, we calculate the detection statistic $\rho$.
The histogram of the detection statistic for $10^5$ of these simulations is provided in Fig.~\ref{fig:yesSignal}. 
The resulting histogram confirms that the distribution for $\Prob(\rho|H_h, h=h^{95\%,\,\text{D.P.}})$ is indeed Gaussian. Moreover, using a Gaussian distribution with the same mean and standard deviation as calculated, we find that $\Prob(\rho>\rho^*|H_h, h=h^{95\%,\,\text{D.P.}}) = 0.946$, almost exactly as expected.

\begin{figure}[ht!]
    \centering
    \includegraphics[width=\linewidth]{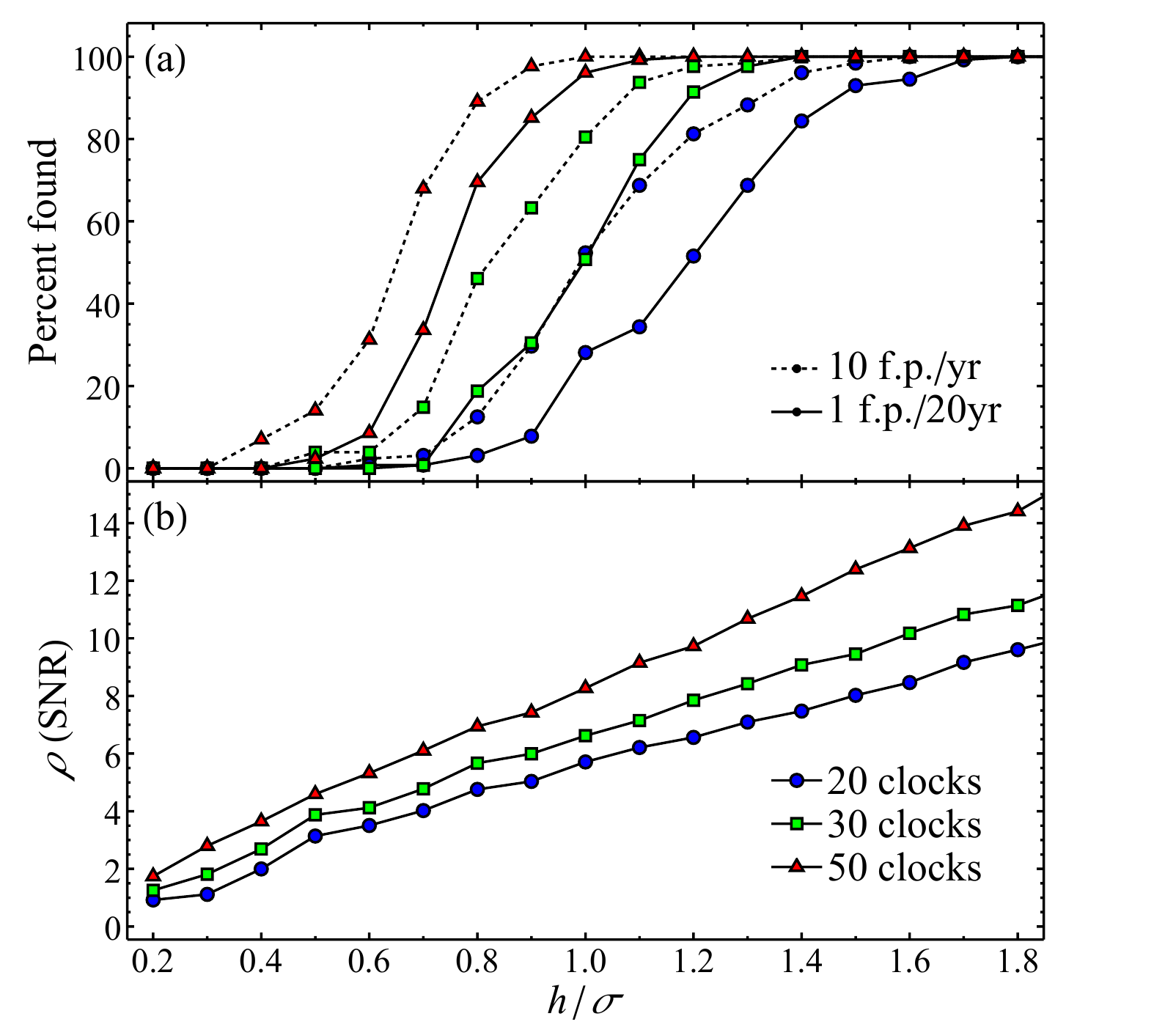}
    \caption{\emph{Top panel:} Detection probability curves as a function of the injected signal strength $h$ for simulated clock networks of varying size. Each point represents the percent of injected events of strength $h$ found for 128 trials. Here $\sigma$ represents the standard deviation of the individual white noise devices that comprise the simulated network. We use $\sigma_\times = 0$ in this simulation. \emph{Bottom panel:} The average detection statistic value $\rho$ of the 128 trials at the time the event was injected.}
    \label{fig:Nscaling}
\end{figure}

A major factor associated with detection probability is the number of devices in the network $\ND$. A more complete discussion of this using the analytic results from Sec.~\ref{sec:TWsnr} is provided in Sec.~\ref{sec:h95dp-scaling}.
Our analysis of detection probability using simulated GPS data was continued by the varying the number of devices in the network $N_D$. We injected signals of varying strength into simulated homogeneous networks of $20, 30,$ and $50$ white noise devices. The percentage of events found as a function of the injected signal strength for these networks is shown in Fig~\ref{fig:Nscaling}. The average value of the detection statistics for each signal strength and clock network is also provided in the same figure. It is clear that our sensitivity to weaker signals improves as the number of devices in the network increases. We have found that $h^{95\%,\,\text{D.P.}} \propto$ $1/\sqrt{\ND}$, as expected.

\begin{figure}[ht!]
    \centering
    \includegraphics[width=\linewidth]{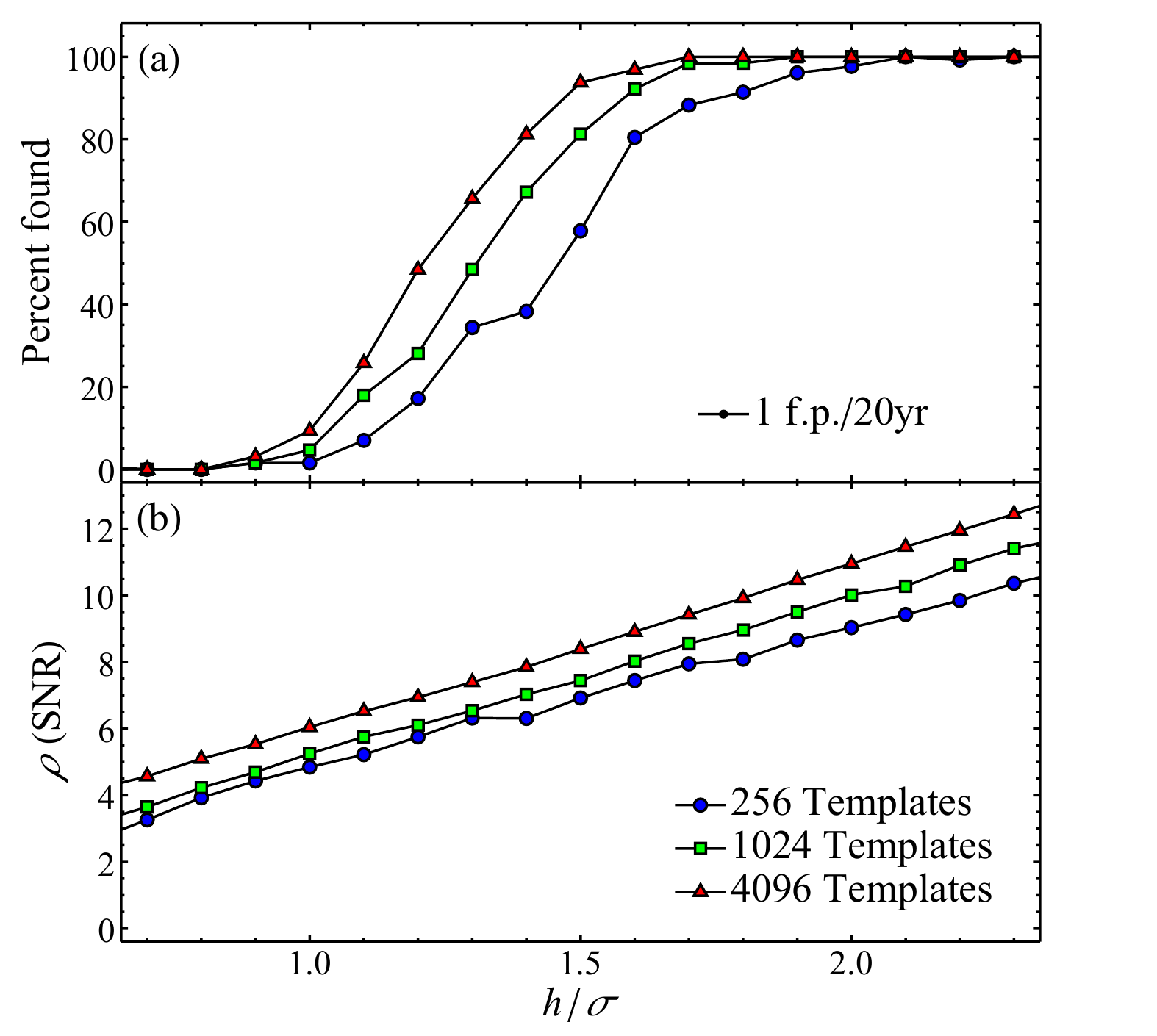}
    \caption{\emph{Top panel:} Detection probability curves as a function of the injected signal strength $h$ for varying template repository sizes. Each point represents the percentage of injected events of strength $h$ found for 128 trials. Here $\sigma$ represents the standard deviation of the individual white noise devices that comprise the simulated network. \emph{Bottom panel:} The average detection statistic value $\rho$ of the 128 trials at the time the event was injected.}
    \label{fig:templateSize}
\end{figure}

To complete our analysis of factors affecting detection probability, we tested the effect of the template repository size $M$. To this end, we simulated a network of 30 homogeneous devices with standard deviation $\sigma=0.05\un{ns}$ and injected events of varying strengths into the data streams. 
The simulated reference clock had a standard deviation of $\sigma_\times = 0.006\un{ns}$.
We then calculated detection statistics for the event using repositories of 256, 1024, and 4096 templates. 
The effect of template size on sensitivity in provided in Fig.~\ref{fig:templateSize} along with the corresponding detection statistic. Notice that an increased template repository size results in better sensitivity and larger SNR values. 
However, increasing the number of templates results in an increase in the false positive rate [Eq.~(\ref{eq:nstar})] along with an increase in computation time. To balance the false positive rate, detection probability and computation time, we typically use 1024 templates in our template repositories.

\section{Parameter estimation}\label{sec:param}
In the event that we find a DM signal in the data stream, our main goal is to estimate the parameters associated with the DM object that caused the signal.
Among the parameters of interest are the incident speed, incident direction, event time, and signal strength. 
The estimates we provide on these parameters correspond to the parameters associated with the model-prescribed signal template that results in an SNR above the detection threshold.

\begin{figure}[ht!]
    \centering
    \includegraphics[width=\linewidth]{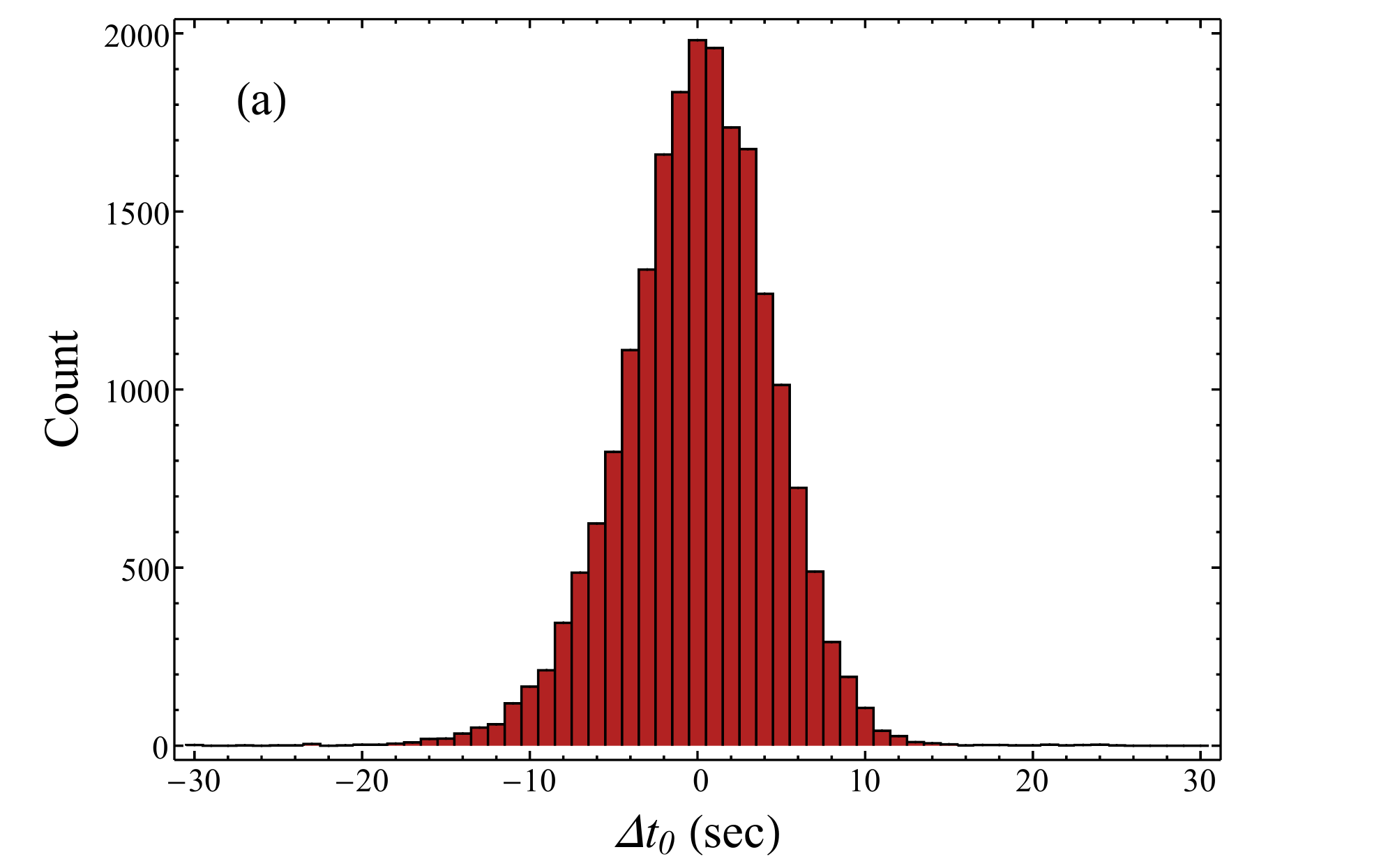}
    \includegraphics[width=\linewidth]{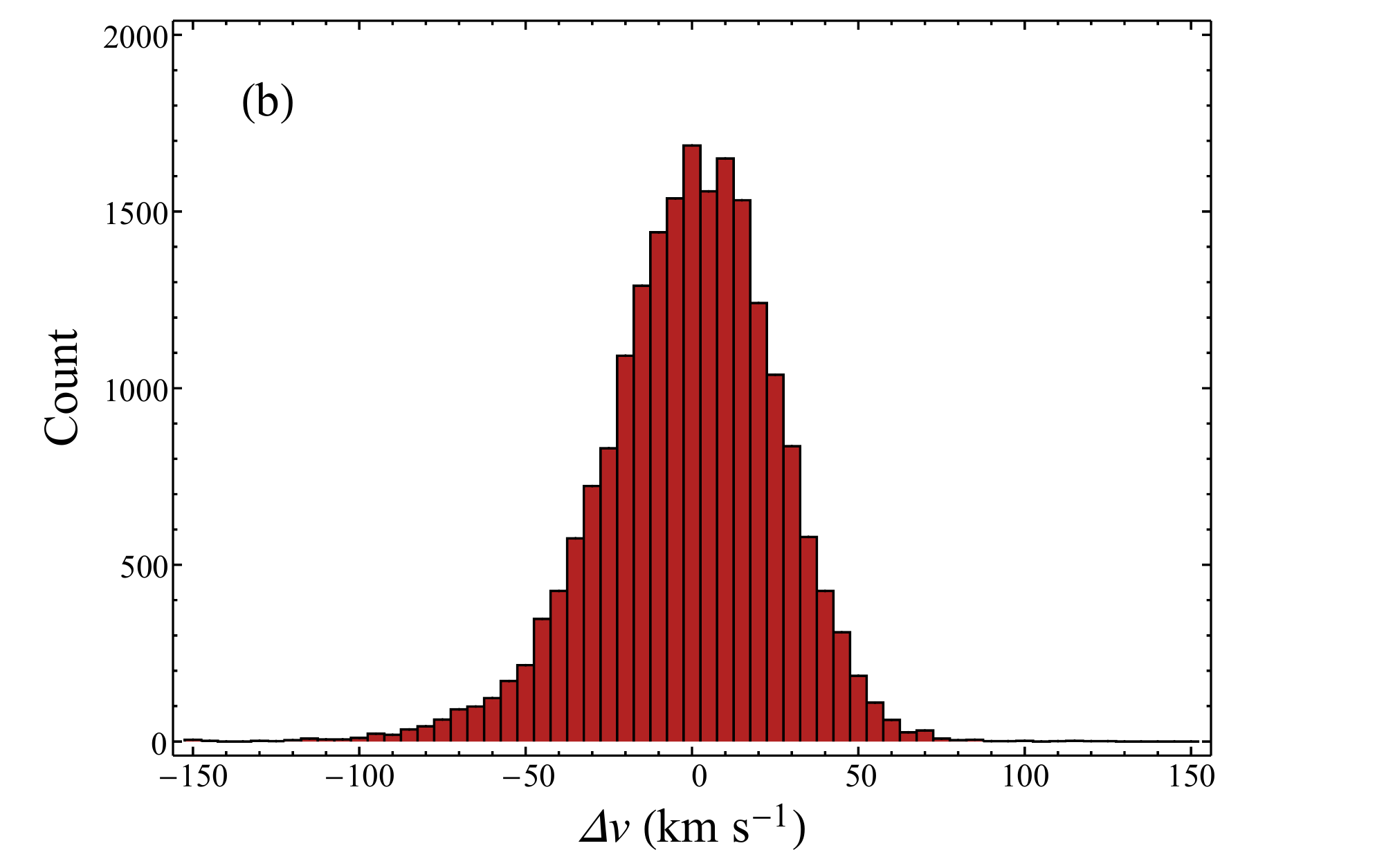}
    \includegraphics[width=\linewidth]{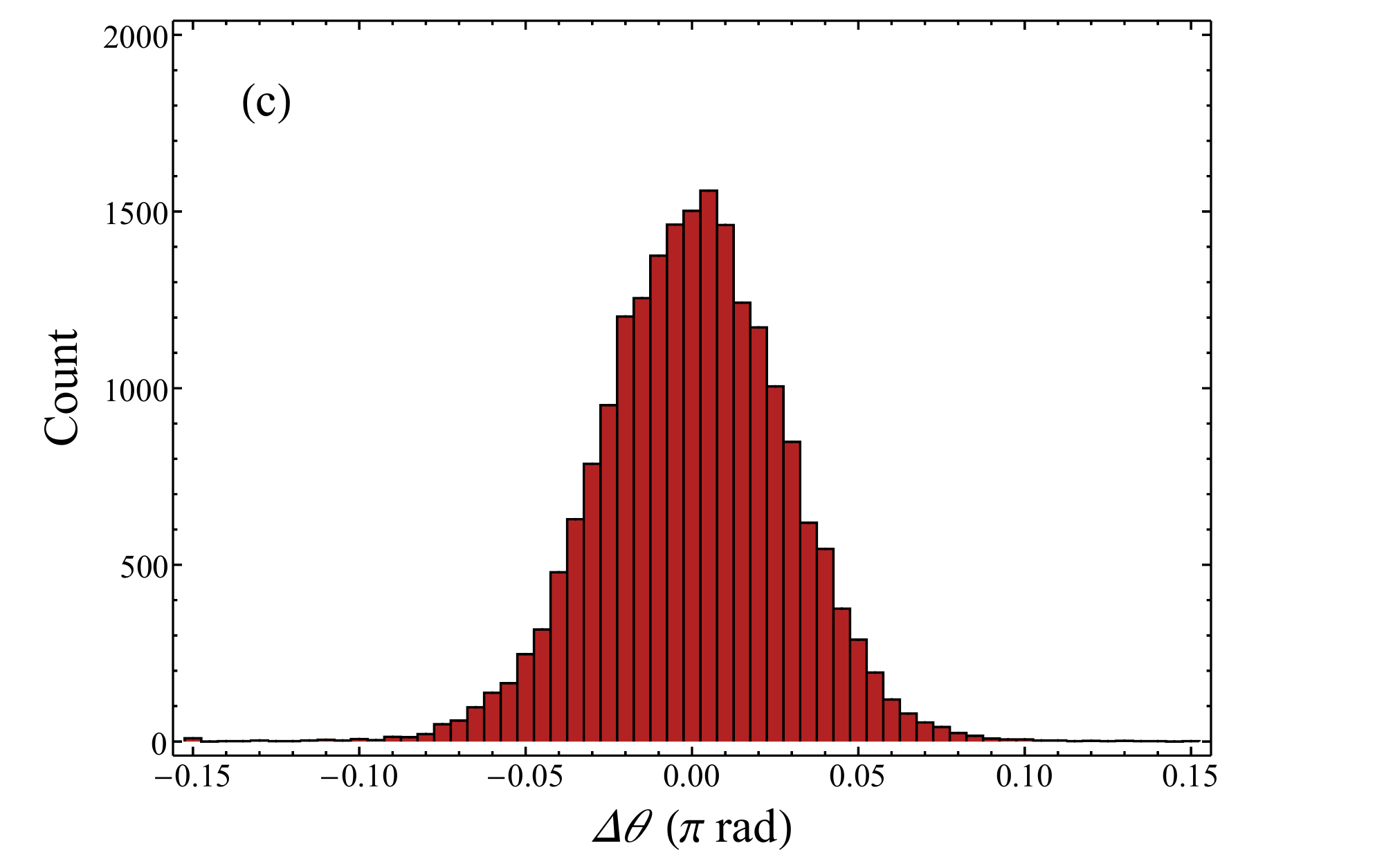}
    \caption{Histograms of the difference in extracted and injected event parameters from parameter estimation routine. \emph{Top}: Event time $t_0$, 1$\sigma$ resolution of $\pm 5\un{s}$. \emph{Middle}: Event velocity $v$, $1\sigma$ resolution of $\pm 27\un{km \, s^{-1}}$. \emph{Bottom}: Incident direction angle $\theta$, resolution of $\pm 0.05\pi$ radians.}
    \label{fig:params}
\end{figure}

In order to test the efficacy of our parameter estimation, we performed $\approx 20,000$ iterations of injecting a DM signal of considerable strength (twice the level of the noise standard deviation, $\sigma = 0.05\un{ns}$) with random parameters into a stream of simulated white-noise data for $N_D=30$ clocks. 
For each iteration, we calculate the SNR for every epoch in the simulated time window and store the event parameters that resulted in an SNR above the detection threshold. 
These extracted parameters are then compared to the injected parameters to check the precision of our parameter extraction. 
Histograms depicting our precision are shown in Fig.~\ref{fig:params}. Our resulting resolution was the following: $\pm 27\un{km \, s^{-1}}$ for velocity, $\pm 0.05\pi$ radians for incident angle $\theta$, and $\pm 5\un{s}$ for the event time.

\section{Placing limits}\label{sec:reach}
\begin{figure*}[ht!]
    \centering
    \includegraphics[width=.49\linewidth]{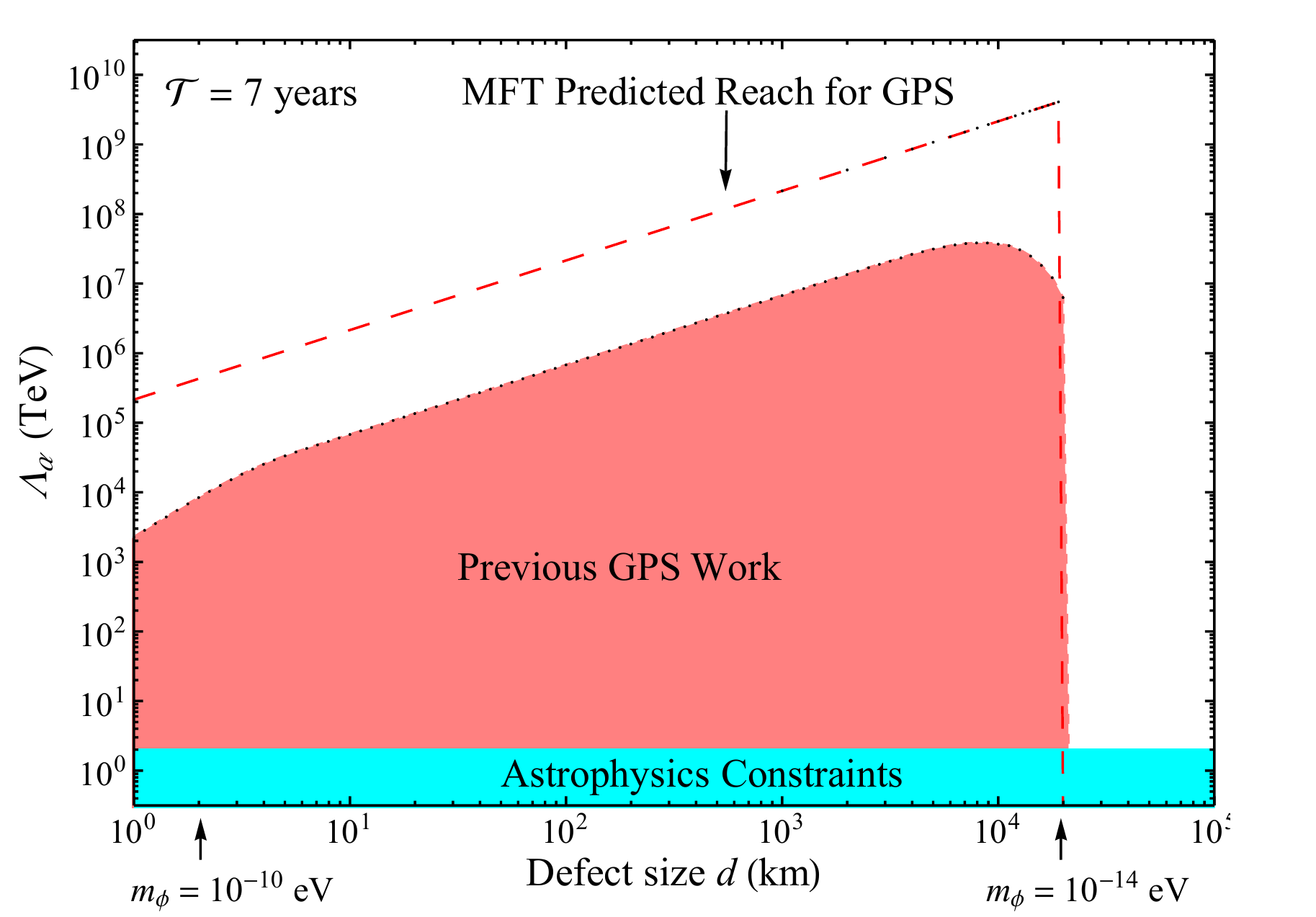}
    \includegraphics[width=.49\linewidth]{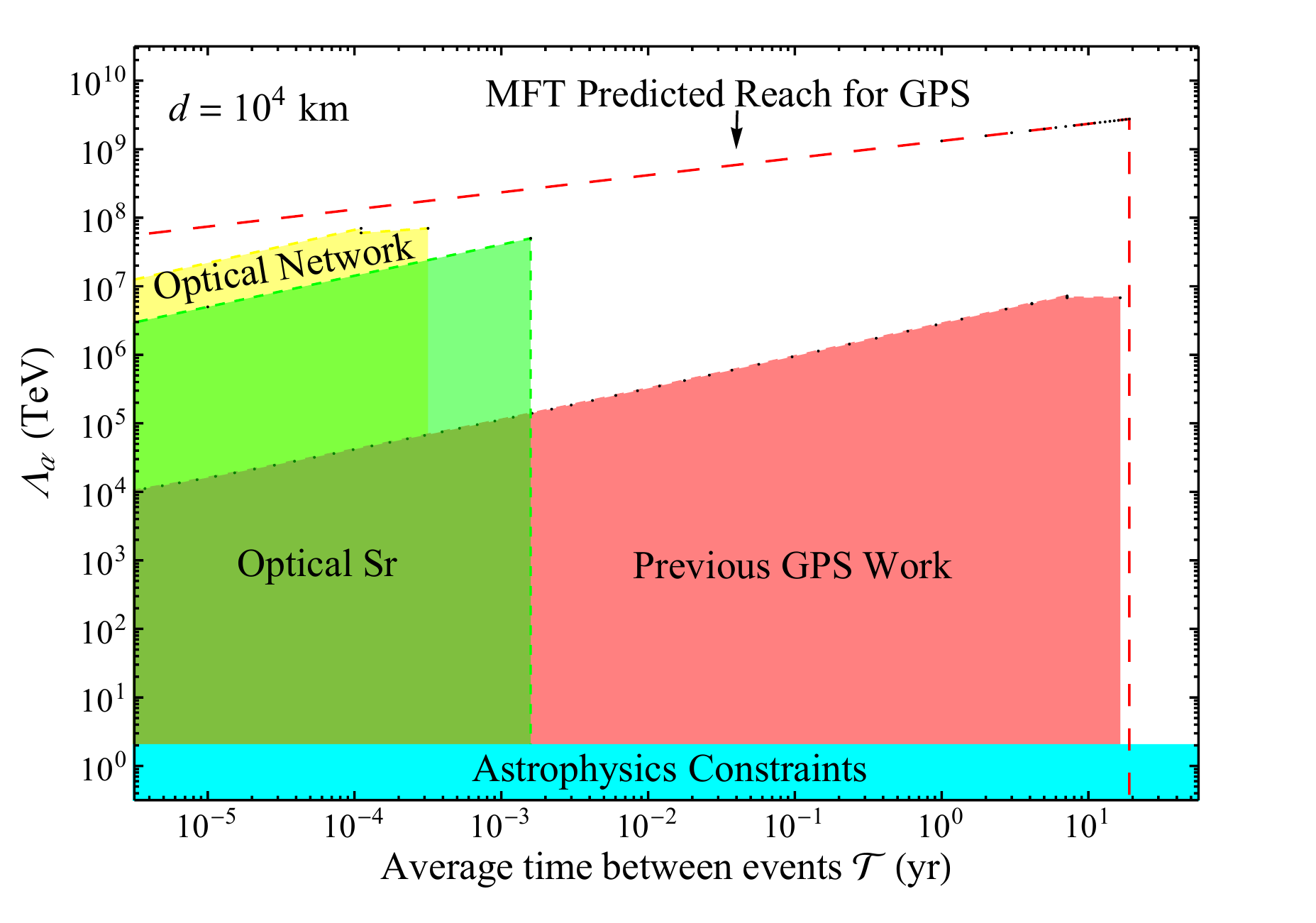}
    \caption{Projected discovery reach for thin wall dark matter objects using the matched-filter technique along with existing constraints. The red dotted lines represent the least stringent and most stringent discovery reaches for the 2010 GPS atomic clock network. The shaded blue regions are the constraints coming from astrophysics~\cite{OLIVE2008} while the salmon shaded regions are the constraints placed by previous work from the GPS.DM collaboration~\cite{Roberts2018b}. The green shaded region contains the constraints placed by optical clock experiments~\cite{Wcislo2016}, while the yellow region contains the constraints from a global terrestrial network of laboratory clocks~\cite{Wcislo2018}.}
    \label{fig:exclusion}
\end{figure*}
Suppose we do not observe a DM interaction signature in the GPS atomic clock data stream. 
This means that there were no SNR values with a magnitude above the detection threshold $\rho^*$. 
We may then establish the lower and upper limits on the DM signal strength $h$.
For the upper limit, suppose the largest SNR value we observed was $\rho_{\text{obs}}$. We define the $95\%$ confidence upper limit $h^{95\%,\,\text{U.L.}}$ as the minimum value of $h$ for which
\begin{equation}
    \Prob(\rho > \rho_{obs}|H_h) = 0.95\, .
\end{equation}
That is, we find the minimum value of $h$ for which we would observe an SNR value larger than $\rho_{obs}$ $95\%$ of the time if there was in fact a signal of strength $h$ in the data stream. 
The $95\%$ confidence lower limit, $h^{95\%,\,\text{L.L.}}$, is defined similarly. Since the SNR $\rho$ is an odd function in $h$,  $h^{95\%,\,\text{L.L.}} = -h^{95\%,\,\text{U.L.}}$.

\subsection{Maximum and minimum sensitivity given clock network characteristics}\label{sec:h95dp-scaling}

Before analyzing the data, we can project a minimum upper limit and maximum lower limit on $h$ by replacing $\rho_{\text{obs}} \to \langle  \rho_{\text{obs}}  \rangle$.
Then, the minimum $95\%$ confidence upper limit for $h$ is the minimum value of $h$ for which
\begin{equation}
     \Prob(\rho >  \langle \rho_{\text{obs}}\rangle \, |h=h^*, H_h) = 0.95\, .
\end{equation}
We will denote the value of $h$ that satisfies this requirement by $h^*$.
Assuming that events are weak, i.e., well below the noise floor, it is clear by the nature of the SNR $ \langle \rho_{\text{obs}}\rangle \to 0$.
Once again, the maximum lower limit is defined in a similar fashion, resulting in $-h^*$ serving as the maximum lower limit for the signal strength.
The maximum possible exclusion limits can be placed on the magnitude of $h$ with $95\%$ confidence by bounding it by $h^*$: $|h|< |h^*|$.

For the idealized sensor network of Sec.~\ref{sec:TWsnr}, we are able to find the exact relation between $h^*$ and the  network characteristics ($\ND, \sigma, \xi,$ and $\eta$) using probability distribution~(\ref{Eq:SNR-prob-distro}).  Ultimately, we find that
\begin{equation}
    h^* = K \frac{\sigma}{\sqrt{\ND N_E}} \sqrt{\frac{1+\xi}{1+\xi+\eta^2}} \, , \label{Eq:hStar}
\end{equation}
where $K$ is determined by the level of confidence (for 95\% confidence $K=1.64$). 
Notice that when $\xi = \ND \sigma_\times^2/\sigma^2  \ll 1$ (i.e., when cross-correlation is negligible) our sensitivity is $K \sigma/\sqrt{\ND N_E (1+\eta^2)}$. 
However, when cross-correlation is considerable, or the network is large ($\xi \gg 1$), the sensitivity becomes $K \sigma/\sqrt{\ND N_E}$. 
Thus, when the reference sensor is noisy, its sensitivity encoded by the constant $\eta = \Geff^R/\Geff^a$ is effectively suppressed.

We can also estimate our minimum sensitivity by projecting a maximum upper limit and minimum lower limit on $h$ by using our detection threshold value as the maximum possible observed SNR value for which we do not claim a detection. 
Then, the minimum 95\% confidence lower limit is given by the minimum value of $h$ for which $ \Prob(\rho > \rho^*|H_h) = 0.95$. It should be clear that the maximum upper limit above is the same as the $95\%$ detection probability signal strength $h^{95\%,\,\text{D.P.}}$.
In this case, $h^{95\%,\,\text{D.P.}}$ scales in the same fashion as $h^*$ above, but the constant of proportionality $K$ increases as a function of the false positive rate and template repository size (for less than 1 false positive in 20 years and 1024 templates in the repository, $K=8.2$).
Ultimately, this makes the minimum sensitivity reach nearly an order of magnitude below the maximum predicted reach.

\subsection{Projected sensitivity and discovery reach}
Given a DM model type (domain wall, monopole, etc.), the signal strength $h$ links to the specific field parameters for those models. 
For a thin domain wall, the average strength $h$ is related to the effective coupling and DM object parameters by 
\begin{equation}\label{eq:havg}
h_{\text{avg}}=\hbar c \Gamma_{\text{eff}} \sqrt{\pi} \rho_{\text{DM}} d^2 \CT \, ,
\end{equation}
in agreement with Ref.~\cite{Roberts2017}.

If we assume that a specific coupling $\Gamma_{X}$ dominates the effective coupling then $\Gamma_{\text{eff}}\to\k_{X}\Gamma_{X}$ in Eq.~(\ref{eq:havg}). 
The limit on $h$, $|h_{\text{avg}}| \leq h^*$ translates into a limit on the coupling constant for a particular coupling to a fundamental constant $\Gamma_X$
\begin{equation}
|\Gamma_{X}| \leq \frac{|h^*|}{\hbar c \rho_{\text{DM}} \CT \sqrt{\pi}d^2 \kappa_{X}} \, ,
\end{equation}
where for idealized network $h^*$ is given by Eq.~(\ref{Eq:hStar}).
This provides projected exclusion limits on the effective energy scale $\Lambda_{X}=1/\sqrt{|\Gamma_{X}|}$,
\begin{equation}
\Lambda_{X}>d\sqrt{\frac{\hbar c \rho_{\text{DM}} \CT \sqrt{\pi} \kappa_{X}}{|h^*|}} \, .
\end{equation}

Our projected discovery reach using $h^*$ for the 2010 GPS network ($\sigma \approx 0.02,\, \sigma_\times \approx 0.006,\, \ND = 33,$ and $\eta \approx 1$), is plotted in Fig.~\ref{fig:exclusion} along with existing constraints. This discovery reach includes the possibility of multiple DM interaction events occurring within the time window of the search, which results in a sensitivity that is comparable to that of optical clocks~\cite{Wcislo2016,Wcislo-clock-network-2018}.
Notice that the projected sensitivity reach in Fig.~\ref{fig:exclusion} exhibits a sharp cutoff for domain-walls of thickness larger than $10^4~\un{km}$ and for average times between events larger than $\CT = 20$ years. 
This is due to the fact that DM objects of size larger than $10^4~\un{km}$ will affect satellite clocks and the reference clock simultaneously, resulting in a no detectable signal is the data stream.
Moreover, for thin domain walls, we require that the signal be present for just a single epoch.
The regime for $d\geq 10^4~\un{km}$ belongs to ``thick'' domain walls (see Ref.~\cite{Roberts2018b}). 
The sharp cutoff for average time between events larger than 20 years comes from the fact that only two decades of archival data exists.

\section{Conclusion}
\label{Sec:Conclusions}

In this paper we focused on detecting dark matter transients with networks of atomic sensors. 
We formalized the desired characteristics of such networks and developed  applications of matched-filter technique in the network settings. We extended the previous literature to the practically important case of a network  with cross-node correlations. This setting is 
especially relevant to GPS atomic clock network. Our simulations have proved the method's signal detection and event parameter estimation capabilities. 

While our paper deals with classical networks of quantum sensors, it is worth noting recent proposals~\cite{KomKesBis14,Komar2016} for massively entangled networks of atomic clocks. In these networks, entanglement is spread not only over an atomic ensemble at a single node, but also over nodes. We leave generalization of our paper to entangled networks for the future work when such entangled networks become a reality.

\acknowledgements
We would like to thank V. Dumont for his contributions on implementing the Cholesky decomposition for covariance matrix inversion.
This work was supported in part by the U.S. National Science Foundation and the office of undergraduate research at the University of Nevada, Reno.
G.P. acknowledges additional support of the McNair Scholars Program.

\appendix

\section{ The network covariance matrix and its perturbative inversion}\label{app:covarInv}

\subsection{ Properties of the network covariance matrix}
The covariance matrix $\bm{E}$ is given by the ensemble average
\begin{equation}
 E_{jl}^{ab}=\av{ n_j^a n_l^b} \equiv E_{(aj)(bl)}  \, , \label{Eq:App:CovarianceDef} 
\end{equation}
where $n_j^a$ is the noise in the datastream of the $a$-th device at the temporal grid point (epoch) $j$, and $\av{n_j^a} = 0$ is assumed. Here subscripts $j$ and $l$ range over epochs and the superscripts $a$ and $b$ span network sensors.  When $a=b$ the covariance refers to a single instrument, while cross-node correlations are given by the $a \neq b$ elements. The matrix $E_{jl}^{ab}$ can be visualized as a 2D matrix with super-indexes $(aj)$ and $(bl)$: $E_{ (aj) (bl)} $.  The dimension of the matrix is determined by the number of devices in the network (excluding the reference clock) and the number of points in data window, $\ND \times J_W$.
Because the data streams are stationary, the covariance matrix only depends on the lag  $|j-l|$. From the definition~(\ref{Eq:App:CovarianceDef}),
it is apparent that the covariance matrix is symmetric with respect to swapping the $(aj)$ and $(bl)$ super-indexes. 
Further, the covariance matrix is positive (semi-) definite. 

For the GPS constellation, as discussed in Sec.~\ref{sec:data}, the noise component entering the definition~(\ref{Eq:App:CovarianceDef})  can be represented as
\begin{equation}\label{eq:App:Noise}
    n_j^a = e_j^a - c_j \, ,
\end{equation}
where $e_j^a$ is the individual clock noise and $c_j$ is the contribution from the reference clock noise common to all data streams.  Then
\begin{align}
 E_{jl}^{ab} = & \langle e_j^a e_l^b \rangle - \langle e_j^a c_l \rangle - \langle c_j e_l^b \rangle +  \langle c_j c_l \rangle \nonumber \\
 = & \langle e_j^a e_l^b \rangle +  \langle c_j c_l \rangle  \, , \label{App:Eq:ExplicitE}
\end{align}
as the reference and the node clock noises are uncorrelated.

While the definition of the covariance matrix~(\ref{Eq:App:CovarianceDef}) explicitly refers to noise, in practice~\cite{Roberts2018b} we use  data $d_j^a$ to compute this matrix (this assumes that DM events are exceedingly rare, so that most of contributions into the above values comes from the intrinsic noise of the network). Notice that in our approximation, the covariance matrix does not depend on the spatial geometry of the network - however it does depend on the network composition. For example, for GPS, if the reference clock is switched to a different clock or a satellite clock is swapped, the covariance matrix is affected and needs to be recomputed.  

To gain an insight into the structure of the covariance matrix, consider a simplifying case: suppose the network is comprised from white-noise devices (including reference clock).
Then
\begin{align}
 E_{jl}^{ab} = \sigma_a^2 \delta_{jl} \delta^{ab} +\sigma_\times^2 \delta_{jl} \, , \label{App:Eq:ExplicitEWN}
\end{align}
where $\sigma_a^2$ and $\sigma_\times^2$ are variances for the individual nodes and the reference clock respectively. The common noise source contributes to all the clocks. 
For example, for $\ND = 2$ nodes and $J_W=3$ time window, the covariance matrix reads
\begin{align}
\bm{E} &=
\left( {\begin{array}{*{20}{c}}
  {\boxed{\begin{array}{*{20}{c}}
  {\sigma _1^2}&0&0 \\ 
  0&{\sigma _1^2}&0 \\ 
  0&0&{\sigma _1^2} 
\end{array}}}& \mbox{\Huge $0$ } \\ 
  \mbox{\Huge $0$ }&{\boxed{\begin{array}{*{20}{c}}
  {\sigma _2^2}&0&0 \\ 
  0&{\sigma _2^2}&0 \\ 
  0&0&{\sigma _2^2} 
\end{array}}} 
\end{array}} \right) \nonumber \\
&\qquad+ \left( {\begin{array}{*{20}{c}}
  {\boxed{\begin{array}{*{20}{c}}
  {\sigma _ \times ^2}&0&0 \\ 
  0&{\sigma _ \times ^2}&0 \\ 
  0&0&{\sigma _ \times ^2} 
\end{array}}}&{\boxed{\begin{array}{*{20}{c}}
  {\sigma _ \times ^2}&0&0 \\ 
  0&{\sigma _ \times ^2}&0 \\ 
  0&0&{\sigma _ \times ^2} 
\end{array}}} \\ 
  {\boxed{\begin{array}{*{20}{c}}
  {\sigma _ \times ^2}&0&0 \\ 
  0&{\sigma _ \times ^2}&0 \\ 
  0&0&{\sigma _ \times ^2} 
\end{array}}}&{\boxed{\begin{array}{*{20}{c}}
  {\sigma _ \times ^2}&0&0 \\ 
  0&{\sigma _ \times ^2}&0 \\ 
  0&0&{\sigma _ \times ^2} 
\end{array}}} 
\end{array}} \right) \, .
\label{App:Eq:Example2x3wn}
\end{align}
The block structure of the covariance matrix is apparent. Each block corresponds to individual sensors and the elements inside each block refer to epochs. 

For colored noise sensors each block is assembled from elements of the auto-correlation functions $A_a(|j-l|)= \langle e_j^a e_l^a \rangle/\sigma_a^2$ and $A_\times(|j-l|) =\langle c_j c_l \rangle/\sigma_\times^2$, so that Eq.~(\ref{App:Eq:ExplicitE}) becomes 
\begin{equation}
    E_{jl}^{ab} = 
 \sigma_a^2 A_a(|j-l|) \delta^{ab} + \sigma_\times^2 A_\times(|j-l|) \, .
 \label{App:Eq:FullCovMatrix}
\end{equation}
Thereby, the covariance matrix is a block matrix: the block diagonals are composed of the sum of cross-node and individual device auto-correlation functions while the off-diagonal blocks contain the cross-node correlation. By the definition of the auto-correlation function, $|A_{a,\times}(|j-l|)| \le 1$, and typically inside each block the elements with larger lag (further away from diagonals) become smaller. $A_{a,\times}(0) = 1$ by definition.

Based on these observations, and  to aid in computer implementation,  we can introduce blocks $\bm{A}^{ab}$ and $\bm{X}^{ab}$, so that the corresponding blocks of the covariance matrix $\bm{E}^{ab} = \bm{A}^{ab} + \bm{X}^{ab} $, with each block internally assembled as (c.f., Eq.~(\ref{App:Eq:FullCovMatrix}))   
\begin{align}
   \left(\bm{A}^{ab} \right)_{ij} &= \sigma_a^2 A_a(|j-i|) \delta^{ab} \, , \\ 
   \left( \bm{X}^{ab} \right)_{ij} &= \sigma_\times^2 A_\times(|j-i|)  \, .
   \label{App:Eq:Blocks}
\end{align}
Because $\bm{X}^{ab}$ does not depend on the particular sensor, we  simply refer to all such blocks as $\bm{\bar{X}}$, i.e., $\bm{X}^{ab} \equiv \bm{\bar{X}}$.

We are interested in the inverse of the covariance matrix  required for computing the SNR statistic~(\ref{eq:gensnr}). 
For our white noise example~(\ref{App:Eq:Example2x3wn}), the inversion of the first  matrix is trivial as it is diagonal. The second (x-node covariance matrix) contribution introduces off-diagonal matrix elements making the inversion difficult. Moreover, the $\times$-node covariance matrix is singular. 

It is instructive to rewrite the definition of the inverse, $\bm{E} \bm{E}^{-1} = \bm{I}$, in our notation, 
\begin{equation}
\sum_{bj}E_{\left(  ai\right)  \left(  bj\right)  }\left(  E^{-1}\right)
_{\left(  bj\right)  \left(  a^{\prime}i^{\prime}\right)  }=\sum_{bj}%
E_{ij}^{ab}\left(  E^{-1}\right)  _{ji^{\prime}}^{ba^{\prime}}=\delta
^{aa^{\prime}}\delta_{ii^{\prime}} \, .
\label{App:Eq:InverseDefinition}
\end{equation}

We derived the covariance matrix inverse  in a closed form for a special case of  white noise devices~(\ref{App:Eq:ExplicitEWN}),  
\begin{equation}
    \Big( E^{-1} \Big)^{ab}_{jl} = \frac{1}{\sigma^2}\delta_{jl}\Big(\delta^{ab}-\frac{1}{\ND} \frac{\xi}{1+\xi} \Big) \,,
    \label{App:Eq:InverseEWN}
\end{equation}
where $\xi \equiv \ND\sigma_{\times}^2/\sigma^2$. 
One can verify directly that Eq.~(\ref{App:Eq:InverseDefinition}) is satisfied.  The inverse retains the same block structure as the original matrix~(\ref{App:Eq:ExplicitEWN}). Its derivation is outlined in the following section.

Additionally, certain simplifications can be obtained using discrete Fourier transformation (DFT) (see, e.g., Ref.~\cite{Derevianko2016a}, where DFT for a network covariance matrix was carried out). In DFT, the transformed matrix becomes block-diagonal, each block being of dimension $\ND$, thus simplifying the inversion procedure. However, this approach also requires DFT of the sought  DM signal. In our work, this signal is non-oscillatory making  the interpretation of the DFT procedure non-transparent; we leave the DFT implementation for future work if the computational speed-up is needed. In this work, the full covariance matrix~(\ref{App:Eq:FullCovMatrix}) is inverted numerically using Cholesky decomposition. Below we outline a perturbative method which holds in the limit when the noise of the reference sensor is well below that of the network sensors. We used this perturbative inversion for older (pre-2015) GPS data, where this approximation remains valid.

\subsection{Perturbative inversion}
This approximation to inverting the network covariance matrix was used in our earlier GPS.DM work~\cite{Roberts2018b} and we will detail it  below.  It relies on the von Neumann series expansion, 
\begin{align}
\left(  \bm{G}+\lambda\bm{F}\right)  ^{-1}&=\bm{G}^{-1}\sum
_{n=0}^{\infty}\left(  -1\right)  ^{n}\lambda^{n}\left(  \bm{F} \bm{G}^{-1}\right)^{n} \nonumber \\
&\approx\bm{G}^{-1}-\lambda\bm{G}^{-1}\bm{F} \bm{G}^{-1} \, ,
\label{App:Eq:PerturbativeExpansionInverse}
\end{align}
where $\bm{G}$ and $\bm{F}$ are matrices and $\lambda$ is the expansion (book-keeping) parameter. This identity can be proven by matching terms of the same power of $\lambda$ in the definition $\left(\bm{G}+\lambda\bm{F}\right)  ^{-1} \left(  \bm{G}+\lambda\bm{F}\right)= \bm{I}$. The series converges as long as the absolute values of eigenvalues of the product matrix $\bm{F} \bm{G}^{-1}$ are smaller than $1/\abs{\lambda}$.

Returning to our covariance matrix $\bm{E}$, we make the following identification using our block decomposition~(\ref{App:Eq:Blocks})
\begin{align}
   \bm{G}^{ab} &= \bm{A}^{aa} \delta^{ab} \, , \\
   \bm{F}^{ab} &= \bar{\bm{X}} \,,
\end{align}
and $\lambda=1$. Notice that 
$\bm{G}$ is a block-diagonal matrix (inverse of such a matrix is again a block-diagonal matrix composed of inverses of original blocks). 

Now we illustrate this perturbative technique for a network of white-noise devices. Here the covariance matrix is given by Eq.~(\ref{App:Eq:ExplicitEWN}). Then the above decomposition leads to
 $  G_{jl}^{ab} = \sigma_a^2 \delta^{ab} \delta_{jl}$, 
 $  F_{jl}^{ab} = \sigma_\times^2 \delta_{jl}$.
Then
\begin{equation}
\left(  E^{-1}\right)  _{ij}^{ab}=\left(  \frac{1}{\sigma_{a}^{2}}\delta
^{ab}-\frac{\sigma_{\times}^{2}}{\sigma_{a}^{2}\sigma_{b}^{2}}\right)
\delta_{ij} \, ,
\end{equation} 
which are the  two leading terms in the expansion of the exact result~(\ref{App:Eq:InverseEWN}) in $\xi$.
Nominally, the contribution of the second, perturbative, term is suppressed when $\sigma_\times \ll \min_a (\sigma_a)$, i.e., this approximation can only be used for networks of sensors that have noise levels far greater than that of the reference sensor.
In cases when the reference sensor noise is comparable to that of the network sensors, the approximation breaks down and either an exact inversion must be used or mitigation techniques must be implemented to eliminate or minimize the reference sensor contribution to the individual sensor data streams (see Sec.~\ref{sec:mitigation}).

For the  general case of colored noise sensors, we return to our block 
decomposition~(\ref{App:Eq:Blocks}) of the network covariance matrix and focus on a single block, 
$$(\bm{E}^{-1})^{ab} \approx (\bm{G}^{-1})^{ab}- 
(\bm{G}^{-1}\bm{F} \bm{G}^{-1})^{ab} \,.$$
Because $\bm{G}$ is block-diagonal,
$
\left( \bm{G}^{-1} \right)^{ab} = \delta^{ab} \left(\bm{A}^{aa}  \right)^{-1}
$.
To simplify the second term in the expansion, recall the product rule for block matrices,
\begin{equation}
  \left( {\bm{A} \bm{B}}\right) ^{ab}=\sum_{c=1}^{\ND} \bm{A}^{ac} \bm{B}^{cb} \, .
\end{equation}
This rule parallels the conventional matrix multiplication with individual matrix elements replaced with blocks. Then
\begin{equation}
(\bm{E}^{-1})^{ab} \approx \delta^{ab} \left(\bm{A}^{aa}  \right)^{-1} -  \left(  \bm{A}^{aa}\right)  ^{-1} \bar{\bm{X}} \left(
\bm{A}^{bb} \right)^{-1} \,.
\end{equation}

Further approximation may consist in neglecting off-diagonal matrix elements of the $\times$-node correlation function in the above expression~\cite{Roberts2018b},
$\bar{\bm{X}} \equiv \sigma_\times^2 \bm{I}$.
In this secondary approximation,
\begin{equation}
(\bm{E}^{-1})^{ab} \approx \delta^{ab} \left(\bm{A}^{aa}  \right)^{-1} - \sigma_\times^2 \left(  \bm{A}^{aa}\right)  ^{-1}\left(
\bm{A}^{bb} \right)^{-1} \,.
\end{equation}
This is the approximation used in our calculations for pre-2015 GPS network generations. We also point out that the exact covariance matrix inversion~(\ref{App:Eq:InverseEWN}) for our idealized network of white noise sensors can be derived by following these  block-matrix steps and summing the von Neumann series~(\ref{App:Eq:PerturbativeExpansionInverse}) to all orders. 

\subsection{Performance comparison between exact and perturbative inversion}
In order to utilize the perturbative inversion outlines in the above section, we require that the reference device noise be sufficiently smaller than that of the node devices. 
To verify the inadequacy of the perturbative inversion for networks with a noisy reference sensor, we simulated signal-free data for a homogeneous network of 30 white noise node devices for various reference device noise levels ($\sigma_\times/\sigma =$ 0, 0.1, 0.5, and 1).
We then calculated standard deviation of $\approx 25,000$ template-specific SNR values for the simulated data streams using both the exact inversion  and the approximate inversion of $\boldsymbol{E}$. 
The results of these simulations are provided in Table~\ref{tab:table3}.
We find that the results using the perturbative inversion are nearly identical to the exact inversion for small levels of cross-correlation ($\sigma_\times/\sigma \leq 0.1$). However, when the noise of the reference sensor is large, the approximate inversion behaves poorly compared to the expected value of $\sigma_\rho$. Note that the deviations from $\sigma_\rho = 1$ in the exact inversion column can be attributed to sampling error.
\begin{table}[ht!]
\caption{\label{tab:table3} Comparison of performance of exact numerical inversion of the covariance matrix and the perturbative inversion for various degrees of cross-correlation with simulated white noise data. The computed values for $\sigma_\rho$ are to be compared with the exact analytic result of $\sigma_\rho =1$, see Sec.~\protect\ref{app:SNRVar}.}
\begin{ruledtabular}
\begin{tabular}{ccc}
 $\sigma_\times/\sigma$ & $\sigma_\rho$ (Exact $\boldsymbol{E}^{-1}$) &  $\sigma_\rho$ ( Approx. $\boldsymbol{E}^{-1}$)  \\
\hline
0.0 & 0.96 & 0.98 \\ 
0.1 & 1.02 & 0.99 \\
0.5 & 0.96 & 8.96 \\
1.0 & 0.97 & 31.27 \\
\end{tabular}
\end{ruledtabular}
\end{table}

\section{Derivation of SNR for idealized  network}\label{app:twsnr-deriv}
Consider a homogeneous network of $N_D$ devices each with Gaussian white noise profiles with zero mean and standard deviation $\sigma$, along with a reference sensor also with a Gaussian white noise profile with a standard deviation of $\sigma_\times$ and also with zero mean. That is,
$$
e_j \sim \textrm{Normal}(0,\sigma^2) \qquad \text{and} \qquad
c_j \sim \textrm{Normal}(0,\sigma_\times^2) \, .
$$
Data that contains a dark matter transient signal will be of the form 
\begin{equation}\label{app:eq:data-terms}
    d_j = e_j - c_j + h \Bar{s}_j \, ,
\end{equation}
where $e_j$ is the node sensor Gaussian noise at epoch $j$ (with variance $\sigma^2$), $c_j$ is the reference sensor Gaussian noise at epoch $j$ (with variance $\sigma_\times^2$), and $\Bar{s}_j$ is the ``unit-ized'' DM signal which is scaled by the DM signal strength $h$ (the strength of the signal felt by the network sensors). 
Since the network is assumed to be homogeneous, $h$ is the same for all  network sensors, though we allow for the possibility of the strength of the DM interaction with the reference device to be different than that of the satellite nodes. 
In this case, the reference sensor experiences a signal of strength $h^R$ from the same event in which the network sensors experience a signal of strength $h$.
Then, the unit signal for a sensor $a$ will be of the form (in the case that $h>0$ and the network sensor interacts with the DM wall prior to the reference device) 
\begin{equation}\label{app:eq:twsignal}
    \bar{\boldsymbol{s}}^a = \big\{0, ..., 0, 1, 0, ... 0, -\eta, 0, ..., 0 \big\} \, .
\end{equation}
where $\eta = h^R/h$.
In order to calculate the detection statistic mean [Eq.~(\ref{eq:twmean})] and its variance [Eq.~(\ref{eq:twvar})], we must utilize the inverse of the covariance matrix from Appendix~\ref{app:covarInv}. This is given by
\begin{equation}
    \Big( E^{-1} \Big)^{ab}_{jl} = \frac{1}{\sigma^2}\delta_{jl}\Big(\delta^{ab}-\frac{1}{\ND}\frac{\xi}{1+\xi} \Big) \,,
\end{equation}
where $\xi = \ND\sigma_\times^2/\sigma^2$.

Recall the definition of the template-specific SNR from Eq.~(\ref{eq:gensnr}), and suppose that the template repository contains the exact signal that lies in the data stream, i.e., $\bar{\boldsymbol{s}}_i = \bar{\boldsymbol{s}}$ for some signal template in the repository. Then, the detection statistic is given by $ \rho =  \boldsymbol{d}^T\boldsymbol{E}^{-1}\bar{\boldsymbol{s}}/\sqrt{\bar{\boldsymbol{s}}^T\boldsymbol{E}^{-1}\bar{\boldsymbol{s}}}$.
Thus, calculating the expectation value of $\rho$ and its variance will consist of calculating $\bar{\boldsymbol{s}}^T\boldsymbol{E}^{-1}\bar{\boldsymbol{s}}$ as well as $\boldsymbol{d}^T\boldsymbol{E}^{-1}\bar{\boldsymbol{s}}$.

Using the thin wall template from Eq.~(\ref{app:eq:twsignal}), the vector-matrix-vector product is computed as the sum
\begin{equation*}
    \bar{\boldsymbol{s}}^T\boldsymbol{E}^{-1}\bar{\boldsymbol{s}} =  \sum_{a,b}^{\ND} \sum_{j,l}^{J_W} \bar{s}_j^a \frac{1}{\sigma^2}\delta_{jl} \Big( \delta^{ab} - \frac{1}{\ND}\frac{\xi}{1+\xi} \Big) \bar{s}_l^b \, ,
\end{equation*}
where $\ND$ is the number of network sensors and $J_W$ is the number of epochs in the given time window. Factoring out $1/\sigma^2$, one can sum over the terms separated by the subtraction independently: (1) sum over $\bar{s}_j^a\delta_{jl}\delta^{ab}\bar{s}_l^b$ and (2) sum over $-\bar{s}_j^a \delta_{jl} \frac{1}{\ND}\frac{\xi}{1+\xi}\bar{s}_l^b$. The sum in (1) is just the sum of the squares of all the signal terms from Eq.~(\ref{app:eq:twsignal}) multiplied by the number of devices. 
Since the signal terms are all zero except at epochs where the satellite clock $a$ and the reference clock $R$ are affected, the sum simplifies immensely. 
Now, to compute the  sum in (2), it must be dissected further. 
The Kronecker delta $\delta_{jl}$ in (2) collapses the sum over $l$ and we split the sum in (2) based on whether the signal terms come from distinct sensors or not
\begin{align}\label{app:eq:sum-diff-clocks}
    -\frac{1}{\ND}&\frac{\xi}{1+\xi}  \sum_{a,b}^{\ND} \sum_{j}^{J_W} \bar{s}_j^a\bar{s}_j^b = \nonumber \\
    &-\frac{1}{\ND}\frac{\xi}{1+\xi} \Big( \sum_{a}^{\ND} \sum_{j}^{J_W} (\bar{s}_j^a)^2 + \sum_{a,b\neq a}^{\ND} \sum_{j}^{J_W} \bar{s}_j^a\bar{s}_j^b \Big) \, .
\end{align}
Notice that the first term on the right side of this equation is the same as (1) above.
Now, since every clock experiences the same signal term at the epoch when the reference sensor interacts with the DM wall (epoch $j_R$), the second term on the right can undergo yet another dissection
\begin{equation}\label{app:eq:sum-diff-clocksj}
    \sum_{a,b\neq a}^{\ND} \sum_{j}^{J_W} \bar{s}_j^a\bar{s}_j^b = \sum_{a,b\neq a}^{\ND} \Big( \bar{s}_{j_R}^a \bar{s}_{j_R}^b + \sum_{j\neq j_R}^{J_W} \bar{s}_j^a\bar{s}_j^b \Big) \, .
\end{equation}
The first term in parentheses on the right side is simply $\eta^2$. 
Since the signal template values at all epochs consists of entirely null values besides the epochs where the individual sensors and the reference sensor interact with the DM wall, the second term on the right is only non-zero when there are disparate sensors that are affected by the DM object at the same epoch. 
We denote the ratio of network sensors that are close enough spatially to interact with the DM wall within the sampling time interval $\tau_0$ as $\lambda$. For GPS, $\tau_0=30 \units{s}$ and at galactic velocities, this ``fractional degeneracy'' factor $\lambda\approx 0.2$.
Ultimately, we arrive at
\begin{align}\label{app:eq:sHs}
    \bar{\boldsymbol{s}}^T\boldsymbol{E}^{-1}\bar{\boldsymbol{s}} &= \frac{1}{\sigma^2} \Big[ (1+\eta^2)\ND \nonumber \\
    &\quad - \frac{\xi}{1+\xi} \Big( (1+\eta^2) + (\ND-1)( \eta^2 + \lambda)\Big) \Big] \, .
\end{align}
For a time window large enough to contain multiple non-overlapping events, it is clear that $\bar{\boldsymbol{s}}^T\boldsymbol{E}^{-1}\bar{\boldsymbol{s}} \to N_E \, \bar{\boldsymbol{s}}^T\boldsymbol{E}^{-1}\bar{\boldsymbol{s}}$, where $N_E$ is the number of events contained within the $J_W$ time window.

To derive $\boldsymbol{d}^T\boldsymbol{E}^{-1}\bar{\boldsymbol{s}}$, we recall that each network sensor data term is the sum of noise (comprised of node sensor noise and reference sensor noise) and a signal [Eq.~(\ref{app:eq:data-terms})]. Then, 
\begin{equation}
    \boldsymbol{d}^T\boldsymbol{E}^{-1}\bar{\boldsymbol{s}} = h\bar{\boldsymbol{s}}^T\boldsymbol{E}^{-1}\bar{\boldsymbol{s}} + \boldsymbol{n}^T\boldsymbol{E}^{-1}\bar{\boldsymbol{s}} \, ,
\end{equation}
where the elements of $\boldsymbol{n}$ are given by Eq.~(\ref{eq:App:Noise}).
Since the elements of $\boldsymbol{n}$ are Gaussian random variables with  zero mean, the quantity $\boldsymbol{n}^T\boldsymbol{E}^{-1}\bar{\boldsymbol{s}}$ will also be Gaussian distributed with a mean of zero. Furthermore, since $\bar{\boldsymbol{s}}^T\boldsymbol{E}^{-1}\bar{\boldsymbol{s}}$ is a constant given by Eq.~(\ref{app:eq:sHs}), we find that $ \boldsymbol{d}^T\boldsymbol{E}^{-1}\bar{\boldsymbol{s}}$ is Gaussian distributed with a mean of $h\bar{\boldsymbol{s}}^T\boldsymbol{E}^{-1}\bar{\boldsymbol{s}}$ and variance equal to the variance of $\boldsymbol{n}^T\boldsymbol{E}^{-1}\bar{\boldsymbol{s}}$. This implies that the nature of the SNR detection statistic from Eq.~(\ref{eq:gensnr}) is Gaussian as well, with the same mean and standard deviation as $\boldsymbol{d}^T\boldsymbol{E}^{-1}\bar{\boldsymbol{s}}$ scaled by $1/\sqrt{\bar{\boldsymbol{s}}^T\boldsymbol{E}^{-1}\bar{\boldsymbol{s}}}$.

Ultimately, using Eq.~(\ref{app:eq:sHs}), the mean of the SNR when a signal is present is given by
\begin{align}
    \mu_\rho &= \frac{h}{\sigma}\sqrt{\frac{\ND(1+\eta^2+\xi)-\xi}{1+\xi}} \nonumber \\ 
    &\approx \frac{h\sqrt{\ND}}{\sigma}\sqrt{\frac{1+\eta^2+\xi}{1+\xi}} \, ,
\end{align}
where we dropped the dependence on the fractional degeneracy factor $\lambda$.
The variance of the SNR, 
\begin{equation}
    \sigma_\rho^2 = \frac{\var\{\boldsymbol{d}^T\boldsymbol{E}^{-1}\bar{\boldsymbol{s}} \}}{\bar{\boldsymbol{s}}^T\boldsymbol{E}^{-1}\bar{\boldsymbol{s}}} 
    \, , \label{App:Eq:SNR}
\end{equation}
is computed in Sec.~\ref{app:SNRVar}, where we prove that $\sigma_\rho^2 =1$ in the most general case; this holds, in particular, for our idealized network.

\subsection{SNR variance}
\label{app:SNRVar}
The goal of this section is to prove that
\[
\mathrm{Var}\{  \boldsymbol{d}^{T}\boldsymbol{E}^{-1}\bar{\boldsymbol{s}}%
\}  =\bar{\boldsymbol{s}}^{T}\boldsymbol{E}^{-1}\bar{\boldsymbol{s}} \, ,
\]
implying that the SNR variance~(\ref{App:Eq:SNR}) is $\sigma^2_\rho =1$. The proof holds regardless of the
nature of the covariance matrix - for example it applies to colored noise
network with arbitrary cross-node correlations.

Explicitly,
\[
\mathrm{Var}\{  \boldsymbol{d}^{T}\boldsymbol{E}^{-1}\bar{\boldsymbol{s}}%
\}  =\left\langle \left(  \boldsymbol{n}^{T}\boldsymbol{E}^{-1}%
\bar{\boldsymbol{s}}\right)  \left(  \boldsymbol{n}^{T}\boldsymbol{E}^{-1}%
\bar{\boldsymbol{s}}\right)  \right\rangle \, ,
\]
where $\boldsymbol{n}$ is the intrinsic noise. To streamline notation, we will
use Greek letters to index combinations $\left(  a,i\right)$, where as in
Sec.~\ref{app:covarInv} the first index ($a$) enumerates the sensors and the second index ($i$)
spans epochs. Then,%
\begin{align*}
& \left\langle \left(  \boldsymbol{n}^{T}\boldsymbol{E}^{-1}\bar{\boldsymbol{s}}%
\right)  \left(  \boldsymbol{n}^{T}\boldsymbol{E}^{-1}\bar{\boldsymbol{s}}\right)
\right\rangle   = \\
 & \left\langle \left(  \sum_{\alpha\alpha^{\prime}}n_{\alpha
}\left( E^{-1}\right)  _{\alpha\alpha^{\prime}}\bar{s}_{\alpha
^{\prime}}\right)  \left(  \sum_{\beta\beta^{\prime}}n_{\beta}\left(
E^{-1}\right)  _{\beta\beta^{\prime}}\bar{s}_{\beta^{\prime}}\right)
\right\rangle \\
& =\sum_{\alpha\alpha^{\prime}}\sum_{\beta\beta^{\prime}}\left(
E^{-1}\right)  _{\alpha\alpha^{\prime}}\bar{s}_{\alpha^{\prime}%
}\left\langle n_{\alpha}n_{\beta}\right\rangle \left(  E%
^{-1}\right)  _{\beta\beta^{\prime}}\bar{s}_{\beta^{\prime}} \, .
\end{align*}
By the definition of the covariance matrix, $\left\langle n_{\alpha}n_{\beta
}\right\rangle =E_{\alpha\beta}$. Further, $\sum_{\beta}E_{\alpha\beta}\left(
E^{-1}\right)  _{\beta\beta^{\prime}}=\delta_{\alpha\beta
^{\prime}}$, which reduces%
\[
\mathrm{Var}\{ \boldsymbol{d}^{T}\boldsymbol{E}^{-1}\bar{\boldsymbol{s}}%
\}  =\sum_{\alpha\alpha^{\prime}}\bar{s}_{\alpha}\left(  E%
^{-1}\right)  _{\alpha\alpha^{\prime}}\bar{s}_{\alpha^{\prime}}=\bar{\boldsymbol{s}}%
^{T}\boldsymbol{E}^{-1}\bar{\boldsymbol{s}} \, ,
\]
as we intended to prove. From Eq.~(\ref{App:Eq:SNR}), it follows that $ \sigma_\rho^2 = 1$.

\section{Inverse transform sampling (importance sampling)}\label{app:ITS}
Consider a prior probability density function on one of the DM model parameters $p(\theta)$ (e.g., the standard halo model velocity distribution). The cumulative distribution function (CDF) for the prior is defined as
\begin{equation}
    C(u) = \int_{-\infty}^u p(\theta) d\theta \, .
\end{equation}
We can then define $g(u) = \theta = C^{-1}(u)$. 
This is particularly useful for sampling from known probability distributions: if $u$ is randomly drawn from a uniform [0:1] distribution, then the $\theta=g(u)$ values will be drawn from the prior $p(\theta)$ distribution.
This has the affect of concentrating the sampled points in the regions where $p(\theta)$ is large (and thus naturally reducing the probability of false-positives where $p(\theta)$ is small).
Thereby, the priors are taken into account implicitly in the template generation procedure.
Note that this just the standard method of inverse transform sampling.

\bibliography{mathched-filter-gpsdm.bbl}

\end{document}